\documentclass[letterpaper,11pt]{article}
\pdfoutput=1

\usepackage[totalwidth=480pt, totalheight=680pt]{geometry}
\usepackage{setspace}

\usepackage{amsmath}
\usepackage{mathtools}
\usepackage{epstopdf}
\usepackage{fancyvrb}

\usepackage[english,greek]{babel}
\usepackage{verbatim}
\usepackage{booktabs}
\usepackage{arydshln}

\usepackage[usenames,dvipsnames]{color}
\usepackage{color}
\usepackage{pstricks,framed}

\usepackage{tensor}
\usepackage{amssymb}
\usepackage{amsthm}
\usepackage{amsfonts}
\usepackage{simplewick}

\usepackage{chngcntr}
\usepackage[retainorgcmds]{IEEEtrantools}

\usepackage[]{mciteplus}
\mciteSetBstSublistMode{n}
\mciteSetMidEndSepPunct{\quad$\bullet$\quad}{.}{\relax}
\usepackage{graphicx}
\usepackage[colorlinks=true, linktoc=page, citecolor=blue, urlcolor=blue]{hyperref}

\counterwithin{equation}{section}
\begin{document}
\selectlanguage{english}
\title{{\color{Blue}\textbf{The Omega-Infinity Limit of Single Spikes}}\\[12pt]} \date{}
\author{\textbf{Minos Axenides}$^1$, \textbf{Emmanuel Floratos}$^{1,2,3}$ \textbf{and Georgios Linardopoulos}$^{1, 2}$\footnote{E-mails: \href{mailto:axenides@inp.demokritos.gr}{axenides@inp.demokritos.gr}, \href{mailto:mflorato@phys.uoa.gr}{mflorato@phys.uoa.gr}, \href{mailto:glinard@inp.demokritos.gr}{glinard@inp.demokritos.gr}.}\\[12pt]
$^1$ Institute of Nuclear and Particle Physics, N.C.S.R., "Demokritos",\\
153 10, Agia Paraskevi, Greece.\\[6pt]
$^2$ Department of Physics, National and Kapodistrian University of Athens,\\
Zografou Campus, 157 84, Athens, Greece.\\[6pt]
$^3$ Department of Physics, Theory Division, CERN, \\ CH-1211 Geneva 23, Switzerland. \\[12pt]}
\maketitle\flushbottom\normalsize
\begin{abstract}
\normalsize{\noindent A new infinite-size limit of strings in $\mathbb{R}\times\text{S}^2$ is presented. The limit is obtained from single spike strings by letting the angular velocity parameter $\omega$ become infinite. We derive the energy-momenta relation of $\omega = \infty$ single spikes as their linear velocity $v \rightarrow 1$ and their angular momentum $\mathcal{J} \rightarrow 1$. Generally, the $v \rightarrow 1$, $\mathcal{J} \rightarrow 1$ limit of single spikes is singular and has to be excluded from the spectrum and be studied separately. We discover that the dispersion relation of omega-infinity single spikes contains logarithms in the limit $\mathcal{J} \rightarrow 1$. This result is somewhat surprising, since the logarithmic behavior in the string spectra is typically associated with their motion in non-compact spaces such as AdS. Omega-infinity single spikes seem to completely cover the surface of the 2-sphere they occupy, so that they may essentially be viewed as some sort of "brany strings". A proof of the sphere-filling property of omega-infinity single spikes is given in the appendix.}
\end{abstract}
\newpage
\tableofcontents
\section[Introduction and Motivation]{Introduction and Motivation \label{Section:Introduction}}
The AdS/CFT correspondence has been revolutionized during the past ten years by the introduction of integrability methods \cite{KristjansenStaudacherTseytlin09, Beisertetal12} that can be used in order to solve the theories on both sides of the correspondence.\footnote{In the large-$N_c$/planar limit ($N_c = \infty$), the string theory is essentially free ($g_s = \infty$) for $\lambda = 4\pi g_s N_c = \text{const}$. AdS$_5$/CFT$_4$ is also thought to be quantum integrable in the planar limit.} As far as a certain class of (long) rotating strings is concerned however,\footnote{These are dual to long operators of $\mathcal{N} = 4$ super Yang-Mills (SYM) at strong 't Hooft coupling $\lambda \rightarrow \infty$.} integrability tools and techniques (such as the ABA, TBA, Y-system, QSC, etc.) have neither been sufficiently developed, nor do they provide the high-loop spectroscopic predictions that they typically yield at either weak coupling (e.g.\ 9-loop Konishi, \cite{MarboeVolin14}) or at strong coupling and small spins (e.g.\ 3-loop Konishi \cite{GromovLevkovichMaslyukSizovValatka14}).\footnote{Even the more recently developed techniques, such as the quantum spectral curve method \cite{GromovKazakovLeurentVolin13}, succeed only when the coupling is weak and at strong coupling when the spin is small.} In this paper we will be dealing with the spectral problem in precisely one of those regimes (namely long strings in $\mathbb{R}\times\text{S}^2$) where the existing spectroscopic machinery is still too complicated to be used for explicit calculations of operator scaling dimensions and string-state energies. \\[6pt]
\begin{figure}
\begin{center}
\includegraphics[scale=.7]{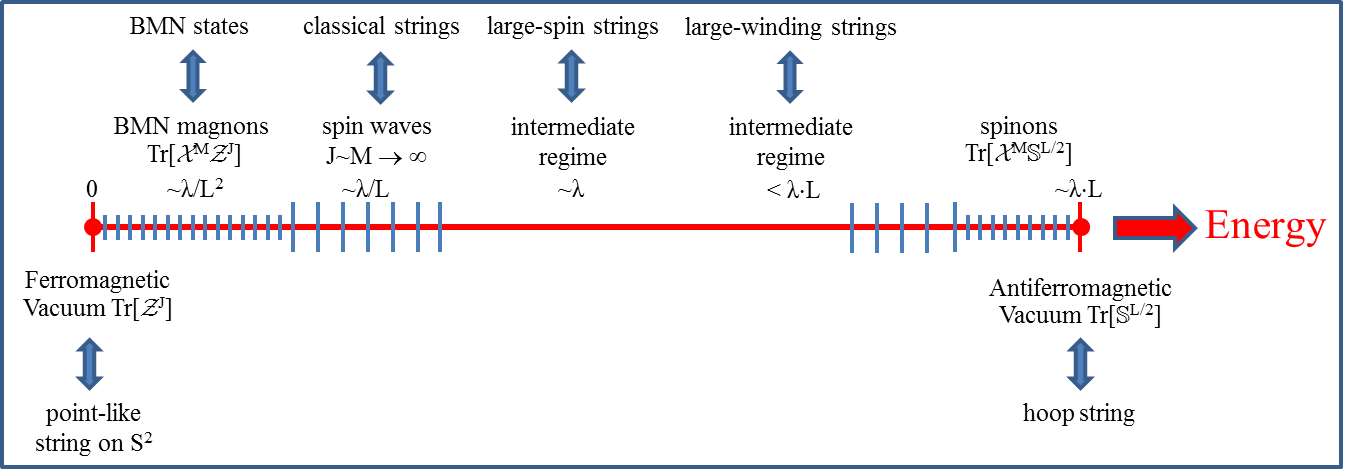}
\caption{Spectrum of AdS$_5$/CFT$_4$. Figure adapted from the talk in \cite{Zarembo08talk}.\label{Graph:AdS5_CFT4_Spectrum}}
\end{center}
\end{figure}
\indent A general description of the spectrum $E - L$ of planar AdS/CFT, based on the structure of the Heisenberg ferromagnet, can be found in the references \cite{RoibanTirziuTseytlin06a} and \cite{Okamura09} (see also figure \ref{Graph:AdS5_CFT4_Spectrum}). Let $E$ denote the scaling dimension and $L$ the length of the $\mathfrak{su}\left(2\right)$ operator.\footnote{In the following, $\mathcal{X}, \ \mathcal{Y}, \ \mathcal{Z}$ will denote the three complex scalar fields of $\mathcal{N} = 4$ SYM, composed out of the theory's six real scalars $\Phi$. It was proven in \cite{MinahanZarembo03} that the one-loop dilatation operator of the $\mathfrak{su}\left(2\right)$ sector of the theory (consisting of all the single-trace operators $\text{Tr}\left[\mathcal{Z}^{J}\mathcal{X}^{M}\right]$) is given by the Hamiltonian of the ferromagnetic XXX$_{1/2}$ Heisenberg spin chain.} We consider the thermodynamic limit $E,L \rightarrow \infty$. Qualitatively the spectrum is expected to be the same for greater values of the 't Hooft coupling $\lambda$. For the strongly coupled description of AdS/CFT, the generic structure of the string spectrum is also expected to be the same. The spectrum will also be similar in all the other sectors of AdS/CFT.\\[6pt]
\indent The bottom of the spectrum is occupied by the BPS operator Tr$\left[\mathcal{Z}^J\right]$ that is the vacuum state of a ferromagnetic spin chain with energy $E - J = 0$. The vacuum is dual to a point-like string that rotates at the equator of S$^2\subset\text{AdS}_5\times\text{S}^5$. Magnons are excitations above the ferromagnetic vacuum with energies $E - J \sim \lambda/J^2$ and large spins, $J \rightarrow \infty$. An $M$-magnon operator Tr$\left[\mathcal{X}^M\mathcal{Z}^J\right]$ ($L = J + M$) is dual to a nearly point-like (BMN) string that rotates in $\mathbb{R}\times\text{S}^2$ with two angular momenta $J_1 = J \rightarrow \infty$ and $J_2 = M$ (finite).\footnote{To get a BMN string \cite{BMN02} one more condition is needed, namely that the ratio $N_c/J^2$ must be held fixed.} For $J \sim M \rightarrow \infty$, we obtain a low-energy spin wave with $E - J \sim \lambda/J$. At the intermediate regime we find states with $E - J \sim \lambda$. The latter are dual to long strings that rotate on the 2-sphere, e.g.\ GKP strings \cite{GubserKlebanovPolyakov02}, giant magnons \cite{HofmanMaldacena06}, etc. \\[6pt]
\indent The spectrum terminates at the antiferromagnetic (AF) vacuum Tr$\left[\mathbb{S}^{L/2}\right]$, where $\mathbb{S}$ are spin-neutral composites of the fields of $\mathcal{N} = 4$ SYM. The AF vacuum has $E - L \sim \lambda\cdot L \rightarrow \infty$ \cite{Zarembo05} and it is dual to the so-called "hoop" string, a circular stationary string of infinite length that is wound around the equator of S$^2$ \cite{Okamura09}. Spinons Tr$\left[\mathcal{X}^M\mathbb{S}^{L - M/2}\right]$, are excitations of the AF state with energies $E - L < \lambda\cdot L \rightarrow \infty$. Of course we have mostly described what happens in the $\mathfrak{su}\left(2\right)$ sector, however one may generalize this discussion to the entire AdS/CFT, by replacing the scalar impurity $\mathcal{X}$ with any of the fields $\left\{\mathcal{Y},F_{\mu\nu},\mathcal{D}_{\mu},\psi_{\text{a},\alpha}\right\}$ of $\mathcal{N} = 4$ SYM, and move from $\mathbb{R}\times\text{S}^2$ to the full AdS$_5\times\text{S}^5$ spacetime in order to get the dual picture. \\[6pt]
\indent The energy of single magnon states Tr$\left[\mathcal{X}\mathcal{Z}^J\right]$ in a spin chain of total length $L = J + 1$ is \cite{BeisertDippelStaudacher04}:
\begin{IEEEeqnarray}{c}
E - J = \sqrt{1 + \frac{\lambda}{\pi^2}\sin^2\left(\frac{p}{2}\right)}, \quad p = \frac{2\pi k}{L},\ k \in \mathbb{Z}, \quad \lambda = g_{\text{YM}}^2 N_c, \label{Magnon1}
\end{IEEEeqnarray}
where $p$ is the magnon's quantized momentum. This formula can also be derived if one extends the corresponding symmetry algebra $\mathfrak{su}\left(2|2\right) \oplus \mathfrak{su}\left(2|2\right) \subset \mathfrak{psu}\left(2,2|4\right)$ by two central charges \cite{Beisert05b}. Extending the symmetry algebra is necessary in order to obtain magnon states with non-vanishing total momentum and violate the cyclicity of the trace condition. For this reason one-magnon operators with non-vanishing momentum $p$ are usually written as states rather than single-trace operators:
\begin{IEEEeqnarray}{c}
\mathcal{O}_M = \sum_{m = 1}^{J+1} e^{imp} \left|\mathcal{Z}^{m-1}\mathcal{X}\mathcal{Z}^{J-m+1}\right\rangle. \label{GM_Operator}
\end{IEEEeqnarray}
\indent \eqref{Magnon1} is only valid asymptotically, namely up to a gauge theory loop-order that is less than the length $L = J + 1$ of the spin chain. When the size of the spin chain becomes infinite ($J = \infty$), \eqref{Magnon1} holds to all loops from weak to strong coupling. In this case we may expand \eqref{Magnon1} at both small and large coupling $\lambda$:
\begin{IEEEeqnarray}{c}
E - J = 1 + \frac{\lambda}{2\pi^2}\sin^2\left(\frac{p}{2}\right) + \ldots \quad \left(\lambda \rightarrow 0\right) \quad \& \quad E - J = \frac{\sqrt{\lambda}}{\pi}\sin\left(\frac{p}{2}\right) + 0 + \ldots \quad \left(\lambda \rightarrow \infty\right). \qquad \label{Magnon2}
\end{IEEEeqnarray}
\indent Now it was shown by Hofman and Maldacena in \cite{HofmanMaldacena06} that it is possible to obtain the string theory duals of magnons in the intermediate energy regime \eqref{Magnon2} (where $E - J \sim \lambda$), by considering a different kind of limit than the BMN. Giant magnons (GM) are classical open strings with a single spin that rotate in $\mathbb{R}\times\text{S}^2 \subset \text{AdS}_5\times\text{S}^5$. They can be identified with the string theory duals of magnons, because their energy-spin relation at tree level
\begin{IEEEeqnarray}{c}
E - J = \frac{\sqrt{\lambda}}{\pi} \, \sin\frac{p}{2}, \quad J = \infty, \ \sqrt{\lambda} = \frac{R^2}{\alpha'} \rightarrow \infty,\footnote{We employ the following convention: with $E,J,p = \infty$ and $v,\omega = 1$ or $\omega = \infty$ we denote infinite size (as obtained by computing the corresponding limits), while with $E,J,p \rightarrow \infty$ and $v,\omega \rightarrow 1$ or $\omega \rightarrow \infty$ we denote large but still finite size.} \label{Magnon3}
\end{IEEEeqnarray}
is identical with the corresponding term in the strong coupling expansion \eqref{Magnon2} of \eqref{Magnon1}. The same is true for the S-matrix of giant magnons, which can be shown to reproduce the result from the gauge theory side. The corresponding infinite-volume one-loop shifts have also been found to agree \cite{PapathanasiouSpradlin07, ChenDoreyLimaMatos07}. \\[6pt]
\indent When the size of the system becomes finite, \eqref{Magnon1} is no longer valid to all loops but there exists a critical loop-order (equal to the length $L$ of the operator) above which \eqref{Magnon1} starts receiving finite-size or wrapping corrections. If the string theory description of strongly coupled $\mathcal{N} = 4$ SYM is correct, we should expect to find these finite-size corrections encoded in the perturbative string expansion even at tree level. Finite-size giant magnons were introduced in \cite{ArutyunovFrolovZamaklar06, AstolfiForiniGrignaniSemenoff07}. A study of their classical finite-size corrections can be found in \cite{FloratosLinardopoulos14}.\footnote{See also the thesis \cite{Linardopoulos15b} and the talk \cite{Linardopoulos15a}.} \\[6pt]
\indent The direct analogs of giant magnons for the AF ground state are single spikes (SSs) \cite{IshizekiKruczenski07, MosaffaSafarzadeh07}. These are the string theory duals of $\mathcal{N} = 4$ SYM (single) spinon states \cite{HayashiOkamuraSuzukiVicedo07},
\begin{IEEEeqnarray}{c}
\mathcal{O}_{\mathbb{S}} \sim \sum_{m = 0}^{(L-1)/2}\left|\mathbb{S}^{m}\,\mathcal{X}\,\mathbb{S}^{(L-1)/2-m}\right\rangle, \quad J \in \mathbb{R}, \quad \lambda,p \rightarrow \infty, \label{SpinonOperators}
\end{IEEEeqnarray}
which are elementary excitations above the AF vacuum state Tr$\left[\mathbb{S}^{L/2}\right]$ \cite{Okamura09}. Finite-size single spikes are open single spin strings that rotate in $\mathbb{R}\times\text{S}^2\subset\text{AdS}_5\times\text{S}^5$ and are described in terms of their linear velocity $v$ and the angular velocity $\omega$:\footnote{The term "angular velocity" is actually a misnomer for the parameter $\omega$, since the true angular velocity of the string is given by $\dot{\phi} = \omega + \dot{\varphi}$, where $\varphi = \varphi\left(\sigma - v\omega\tau\right)$ is a function of the string variables $\sigma$ and $\tau$. The authors kindly apologise to the reader for taking the permission to conventionally call the parameter $\omega$ angular velocity.}
\begin{IEEEeqnarray}{l}
0 \leq 1/\omega \leq \left|v\right| \leq 1 \qquad \text{or} \qquad 0 \leq 1/\omega \leq 1 \leq \left|v\right|.
\end{IEEEeqnarray}
The former regime is known as the elementary region and the latter as the doubled region of single spikes. Like giant magnons, single spikes also belong to the class of long strings/strongly coupled operators for which there are no advanced spectroscopic tools at our disposal. Single spikes can be transformed to giant magnons by means of the $\tau \leftrightarrow \sigma$ transform or "2D duality" \cite{HayashiOkamuraSuzukiVicedo07}:
\begin{IEEEeqnarray}{ll}
\tau \leftrightarrow \sigma \,,\, v \leftrightarrow \frac{1}{\omega} \,,\, \psi \leftrightarrow \left[\frac{\pi}{2} - \psi\right] \quad \Rightarrow \quad \text{Giant Magnons} \leftrightarrow \text{Single Spikes}. \label{TauSigmaTransform}
\end{IEEEeqnarray}
In fact, \eqref{TauSigmaTransform} maps the elementary region of single spikes to the elementary region of giant magnons and the doubled region of single spikes to the doubled region of giant magnons. In \cite{FloratosLinardopoulos14}, two more symmetries between the equations of motion of giant magnons and single spikes were found.\footnote{Namely the $\eta \leftrightarrow -\eta$ and the $\eta \leftrightarrow 1/\eta$ transforms, where $\eta$ is given by equations \eqref{EtaDefinition1}--\eqref{EtaDefinition2}. These either transform between the elementary regions of GMs and SSs, or map their elementary regions to their doubled regions. For more, refer to the appendix A.5 of \cite{FloratosLinardopoulos14}.} \\[6pt]
\indent The dispersion relation of single spikes provides their conserved energy in terms of their other conserved charges, i.e.\ their linear momentum $p$ and their angular momentum $J$. For $v = 1$, the energy and the linear momentum become infinite and single spikes have infinite length and winding (since their length/winding is proportional to their linear momentum). However the difference of the two (scaled) charges is finite:
\begin{IEEEeqnarray}{c}
\mathcal{E} - \frac{p}{2} = \arcsin\mathcal{J}, \quad \mathcal{E} \equiv \frac{\pi E}{\sqrt{\lambda}}, \quad \mathcal{J} \equiv \frac{\pi J}{\sqrt{\lambda}}, \quad p = +\infty, \quad \lambda = \infty. \label{SingleSpike1}
\end{IEEEeqnarray}
For $v = \omega = 1$ the single spike reduces to the infinitely wound hoop string that is dual to the antiferromagnetic ground state Tr$\left[\mathbb{S}^{L/2}\right]$ of $\mathcal{N} = 4$ SYM and has the following dispersion relation:
\begin{IEEEeqnarray}{c}
\mathcal{E} = \frac{p}{2}, \qquad p = +\infty.
\end{IEEEeqnarray}
\indent When $v \rightarrow 1^{\pm}$, single spikes have finite but still large size/winding\footnote{As a reminder, a string of infinite size is defined as having infinite worldsheet size $r$ and therefore infinite energy $E$ (c.f.\ equation \eqref{SS_Energy1} below). Infinite momentum $p$ implies that the angular extent $\Delta\phi$ of the string (or its winding in units of $2\pi$) is also infinite (c.f.\ equation \eqref{BoundaryConditions} below). In our manuscript "string length" is an alias for "string winding", or the length of the string in the target space. The size (i.e.\ the energy) of single spike strings is large/infinite because their energy is equal to their momentum (i.e.\ their length/winding/angular extent) to lowest order (c.f.\ equation \eqref{SingleSpike1}) and the latter is also large/infinite. That is why the terms infinite size/winding are used interchangeably throughout our paper.} and their dispersion relation assumes the following general form \cite{FloratosLinardopoulos14} (at strong coupling, $\lambda = \infty$): \\
\begin{IEEEeqnarray}{l}
\mathcal{E} - \frac{p}{2} = \frac{q}{2} + \sum_{n = 1}^{\infty} \bigg[\hat{\mathcal{A}}_{n0} \, p^{2n - 2} + \ldots + \hat{\mathcal{A}}_{n(2n-2)}\bigg] e^{-n\mathcal{R}} = \frac{q}{2} + \sum_{n = 1}^{\infty} \sum_{m = 0}^{2n - 2} \hat{\mathcal{A}}_{nm}\left(q\right) p^{2n - m - 2} e^{-n\mathcal{R}}, \qquad \label{ClassicalCorrections1}
\end{IEEEeqnarray} \\
where $\hat{\mathcal{A}}_{nm}\left(q\right)$ are some trigonometric coefficients and
\begin{IEEEeqnarray}{ll}
\frac{q}{2} \equiv \arcsin\mathcal{J}, \quad \mathcal{R} \equiv \sqrt{\frac{1}{\mathcal{J}^2} - 1} \cdot \left(p + 2\arcsin\mathcal{J}\right) = \left(p + q\right)\cdot\cot\frac{q}{2}. \qquad
\end{IEEEeqnarray}
In the elementary region ($0 \leq 1/\omega \leq \left|v\right| \leq 1$), the first few terms of \eqref{ClassicalCorrections1} are: \\
\footnotesize\begin{IEEEeqnarray}{ll}
\mathcal{E} = \frac{p}{2} + \frac{q}{2} &{\color{red}+} 4\sin^2\frac{q}{2}\tan\frac{q}{2}\cdot e^{-\mathcal{R}} + \Bigg\{8p^2\cos^2\frac{q}{2} + 2p\cos\frac{q}{2} \left(8q\cos\frac{q}{2} - \sin\frac{3q}{2} + 7\sin\frac{q}{2}\right) + 8q^2\cos^2\frac{q}{2} - 2q\sin q \big(\cos q - \nonumber \\[12pt]
& - 3\big) + \sin^2\frac{q}{2}\left(\cos2q -{\color{red} 2\cos q + 5}\right)\Bigg\}\sec^2\frac{q}{2}\tan\frac{q}{2} \cdot e^{-2\mathcal{R}} + \ldots, \qquad v \rightarrow 1^-, \label{ClassicalCorrections2}
\end{IEEEeqnarray} \normalsize \\
while in the doubled region ($0 \leq 1/\omega \leq 1 \leq \left|v\right|$), the corresponding coefficients are given by: \\
\begin{figure}
\begin{center}
\includegraphics[scale=.7]{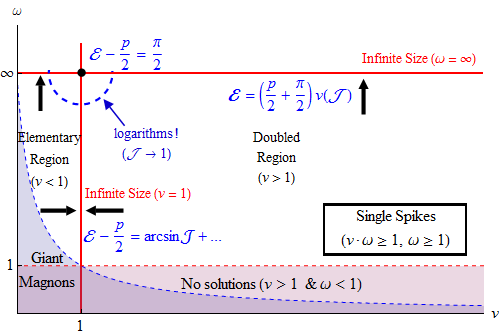}
\caption{Regions \& infinite-size limits of elementary/doubled single spikes.} \label{Graph:SingleSpikeRegions1}
\end{center}
\end{figure}
\footnotesize\begin{IEEEeqnarray}{ll}
\mathcal{E} = \frac{p}{2} + \frac{q}{2} &{\color{red}-} 4\sin^2\frac{q}{2}\tan\frac{q}{2}\cdot e^{-\mathcal{R}} + \Bigg\{8p^2\cos^2\frac{q}{2} + 2p\cos\frac{q}{2} \left(8q\cos\frac{q}{2} - \sin\frac{3q}{2} + 7\sin\frac{q}{2}\right) + 8q^2\cos^2\frac{q}{2} - 2q\sin q \big(\cos q - \nonumber \\[12pt]
& - 3\big) + \sin^2\frac{q}{2}\left(\cos2q -{\color{red} 34\cos q - 91 + 64\csc^2\frac{q}{2}}\right)\Bigg\}\sec^2\frac{q}{2}\tan\frac{q}{2} \cdot e^{-2\mathcal{R}} - \ldots, \qquad v \rightarrow 1^+. \label{ClassicalCorrections3}
\end{IEEEeqnarray} \normalsize
More terms can be found in appendix B of the paper \cite{FloratosLinardopoulos14}. For the $\mathsf{Mathematica}$ code that generates these expansions, the reader may refer to the thesis \cite{Linardopoulos15b}. The above results are in complete agreement with the $v = 1$ infinite-size/winding limit \eqref{SingleSpike1} that we saw above. Just note that if we set $v = 1$, the momentum becomes infinite ($p = +\infty$) and all the exponentials in \eqref{ClassicalCorrections1}, \eqref{ClassicalCorrections2} and \eqref{ClassicalCorrections3} vanish. \eqref{ClassicalCorrections2} also agrees with the leading finite-size correction computed by Ahn and Bozhilov in \cite{AhnBozhilov08a}. \\[6pt]
\indent In the limit $\mathcal{J} = 1 \Leftrightarrow q = \pi$, it seems that the exponential suppression drops out of \eqref{ClassicalCorrections1} since $\mathcal{R} = 0$. A closer look at the expansions \eqref{ClassicalCorrections2}--\eqref{ClassicalCorrections3} reveals that many of the trigonometric coefficients $\hat{\mathcal{A}}_{nm}\left(q\right)$ blow up in the limit $q = \pi$, signaling that the value $\mathcal{J} = 1 \Leftrightarrow q = \pi$ (in the large-winding region $v \rightarrow 1^{\pm} \Rightarrow p \rightarrow +\infty$) is probably singular and has to be treated separately. Furthermore, for $\omega = \infty$ it is possible that the spin $\mathcal{J}$ approaches unity from above ($\mathcal{J} \rightarrow 1^+$---see figure \ref{Graph:InfiniteSizeLimits}) and $\mathcal{R}$ essentially becomes complex. Indeed, it was proven in \cite{FloratosLinardopoulos14} that the algorithm leading to \eqref{ClassicalCorrections2}--\eqref{ClassicalCorrections3} is dysfunctional for $\omega = \infty$, $v \rightarrow 1^{\pm} \Leftrightarrow \mathcal{J} \rightarrow 1^{\pm}$, and it cannot be used. \\[6pt]
\indent The aim of the present paper is to clarify the above situation by computing the dispersion relation of classical single spikes in the infinite-size/winding limit (see figure \ref{Graph:SingleSpikeRegions1}):
\begin{IEEEeqnarray}{ll}
p = +\infty, \quad \mathcal{J} \rightarrow 1^{\pm} \quad \Leftrightarrow \quad \omega = \infty \quad \& \quad v \rightarrow 1^{\pm}, \label{LargeWindingLimit1}
\end{IEEEeqnarray}
where \eqref{ClassicalCorrections1} is no longer valid. \\[6pt]
\indent To get going, let us first note that the $\omega = \infty$ limit of single spikes is an infinite-size/momentum limit, that is
\begin{IEEEeqnarray}{ll}
\omega = \infty \quad \Rightarrow \quad \mathcal{E} = \infty \quad \& \quad p = +\infty,
\end{IEEEeqnarray}
a fact already anticipated in \cite{FloratosLinardopoulos14}. The dispersion relation of single spikes in the $\omega = \infty$ limit becomes:
\begin{IEEEeqnarray}{ll}
\mathcal{E} = \left(\frac{p}{2} + \frac{\pi}{2}\right) \cdot v\left(\mathcal{J}\right), \qquad p = +\infty, \quad \lambda = \infty, \label{SingleSpike2}
\end{IEEEeqnarray}
where $v\left(\mathcal{J}\right)$ is a function of the (scaled) angular momentum $\mathcal{J}$ that depends on the region (elementary/doubled) of single spikes that we are considering. In figure \ref{Graph:InfiniteSizeLimits}, $v\left(\mathcal{J}\right)$ corresponds to the purple line that passes through the point $v\left(\mathcal{J} = 1\right) = 1$. Interestingly, \eqref{SingleSpike2} agrees with \eqref{SingleSpike1} in the double infinite-size/winding limit $\omega = \infty$, $v = 1$:
\begin{figure}
\begin{center}
\includegraphics[scale=0.6]{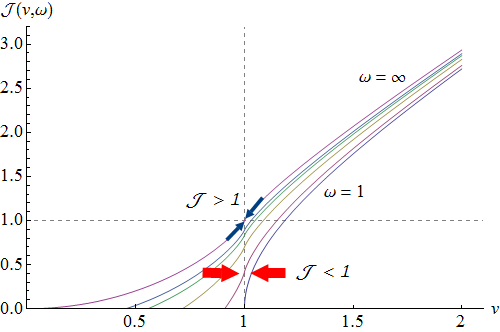}
\caption{Spin of single spikes versus their velocity $v$ for various angular momenta $\omega$.} \label{Graph:InfiniteSizeLimits}
\end{center}
\end{figure}
\begin{IEEEeqnarray}{ll}
\mathcal{E} = \frac{p}{2} + \frac{\pi}{2}, \qquad p = +\infty, \quad \mathcal{J} = 1, \quad \lambda = \infty. \label{SingleSpike3}
\end{IEEEeqnarray}
As we have already said, we have computed \eqref{SingleSpike2} in the region \eqref{LargeWindingLimit1} of single spikes. If we define,
\begin{IEEEeqnarray}{ll}
v\left(\mathcal{J}\right) = \left[1 - \chi\left(\mathcal{J}\right)\right]^{\pm 1/2} \rightarrow 1^{\mp}, \quad \mathcal{L} \equiv \left|1 - \mathcal{J}\right| \rightarrow 0^+, \quad \mathcal{S} \equiv -\ln\mathcal{L} \rightarrow +\infty,
\end{IEEEeqnarray}
where the upper signs refer to the elementary region and the lower signs to the doubled region of single spikes, then the general form of $\chi\left(\mathcal{J}\right)$ in the region \eqref{LargeWindingLimit1} is the following:
\begin{IEEEeqnarray}{ll}
\chi\left(\mathcal{J}\right) &= \frac{4\mathcal{L}}{\mathcal{S}}\cdot\left\{\sum_{n = 1}^{\infty}\frac{\ln^n\mathcal{S}}{\mathcal{S}^{n}} + \sum_{n = 2}^{\infty}\rho_{(nn-1)}\frac{\ln^{n-1}\mathcal{S}}{\mathcal{S}^{n}} + \sum_{n = 3}^{\infty}\rho_{(nn-2)}\frac{\ln^{n-2}\mathcal{S}}{\mathcal{S}^{n}} + \ldots + 1 + \sum_{n = 1}^{\infty}\frac{\rho_{n}}{\mathcal{S}^n}\right\} =\footnote{Following the notation that is used for twist-2 operators (or equivalently for GKP strings in AdS$_3$) \cite{FloratosGeorgiouLinardopoulos13}, we have defined $\rho_{(n0)} \equiv \rho_{n}$, for $n = 0,1,2, \ldots$} \nonumber \\[12pt]
&= \frac{4\mathcal{L}}{\mathcal{S}}\cdot\sum_{n = 0}^{\infty}\sum_{k = 0}^{n} \rho_{(nk)}\left(\mathcal{L}\right) \, \frac{\ln^{k}\mathcal{S}}{\mathcal{S}^{n}}, \label{InverseSpinFunction1}
\end{IEEEeqnarray} \\
where the $\rho_{(nk)}$'s are given by
\begin{IEEEeqnarray}{ll}
\rho_{(nk)}\left(\mathcal{L}\right) = \sum_{m = 0}^{n - k} \rho_{(nkm)} \cdot \mathcal{L}^{n - k - m}.
\end{IEEEeqnarray}
\indent The proof of the dispersion relation \eqref{InverseSpinFunction1}, along with the computation of some of the coefficients $\rho_{(nk)}$ in both the elementary and the doubled region are the main results of our paper. The structure of \eqref{InverseSpinFunction1} for infinite size/winding single spikes is highly reminiscent of the large-spin expansions of the anomalous dimensions of twist-2 operators and those of Gubser-Klebanov-Polyakov (GKP) strings in AdS$_3$ \cite{FloratosGeorgiouLinardopoulos13}. However, the leading logarithm $\ln\mathcal{S}$, as well as all the leading terms $\ln^n\mathcal{S}/\mathcal{S}^n$ that were present in the dispersion relation of twist-2 operators and GKP strings, are missing from \eqref{InverseSpinFunction1}. Yet another difference is that the coefficients $\rho_{(nk)}$ of \eqref{InverseSpinFunction1} depend explicitly on the spin $\mathcal{J}$. \\[6pt]
\indent Our paper is organized as follows. In \S\ref{Section:SingleSpikes} we briefly revisit the definition of finite-size single spikes. In \S\ref{Section:OmegaInfinityLimit} we study the dispersion relation of single spikes in the limit $\omega = \infty$, in both the elementary and the doubled region. We discuss our results in \S\ref{Section:Discussion}. In appendix \ref{Appendix:SolutionsCharges} we establish the consistency of the $\omega = \infty$ limit of single spikes by explicitly deriving the corresponding string sigma model solutions and conserved charges. We also provide a proof of the space-filling property of omega-infinity single spikes.
\section[Finite-Size Single Spikes]{Finite-Size Single Spikes \label{Section:SingleSpikes}}
As we have already mentioned, single spikes are open single spin strings that rotate inside the subset $\mathbb{R}\times\text{S}^2 \subset \text{AdS}\times\text{S}^5$. Let us see how they can be obtained from the generic configuration of open (bosonic) strings in $\mathbb{R}\times\text{S}^2$:
\begin{IEEEeqnarray}{c}
\Big\{t = t\left(\tau,\sigma\right), \, \rho = \overline{\theta} = \overline{\phi}_1 = \overline{\phi}_2 = 0\Big\} \times \Big\{\theta = \theta\left(\tau,\sigma\right), \, \phi = \phi\left(\tau,\sigma\right), \, \theta_1 = \phi_1 = \phi_2 = 0\Big\}. \qquad \label{GM_Ansatz1}
\end{IEEEeqnarray}
In this paper we work with the following parametrization of AdS$_5 \times \text{S}^5$:
\begin{IEEEeqnarray}{ll}
ds^2 = R^2 \Big[-\cosh^2\rho \, dt^2 &+ d\rho^2 + \sinh^2\rho \, \Big(d\overline{\theta}^2 + \sin^2\overline{\theta} \, d\overline{\phi}_1^2 + \cos^2\overline{\theta} \, d\overline{\phi}_2^2\Big) + \nonumber \\[6pt]
& + d\theta^2 + \sin^2\theta \, d\phi^2 + \cos^2\theta \, \left(d\theta_1^2 + \sin^2\theta_1 \, d\phi_1^2 + \cos^2\theta_1 \, d\phi_2^2\right)\Big] \qquad
\end{IEEEeqnarray}
and set
\begin{IEEEeqnarray}{ll}
z\left(\tau,\sigma\right) = R\,\cos\theta\left(\tau,\sigma\right).
\end{IEEEeqnarray}
In the above parametrization, the string's embedding coordinates become:
\begin{IEEEeqnarray}{ll}
Y_0 + i \, Y_5 = R \, e^{i\,t\left(\tau,\sigma\right)} \quad \& \quad & X_1 + i X_2 = \sqrt{R^2 - z^2\left(\tau,\sigma\right)} \cdot e^{i \, \phi\left(\tau,\sigma\right)}, \quad z \in \left[-R,R\right], \quad \phi \in \left[0,2\pi\right) \nonumber \\[6pt]
& X_3 = z\left(\tau,\sigma\right),
\end{IEEEeqnarray}
while the conformal ($\gamma_{ab} = \eta_{ab}$) string Polyakov action is given by
\begin{IEEEeqnarray}{l}
\mathcal{S}_P = \frac{\sqrt{\lambda}}{4\pi} \int d\tau d\sigma \Bigg\{-\left(\dot{t}^2 - t'^2\right) + \frac{\dot{z}^2 - z'^2}{R^2 - z^2} + \frac{1}{R^2}\left(R^2 - z^2\right)\left(\dot{\phi}^2 - \phi'^2\right)\Bigg\}. \label{GMAction1}
\end{IEEEeqnarray}
In the static time gauge ($t = \tau$) the corresponding Virasoro constraints and Pohlmeyer reduction become:
\begin{IEEEeqnarray}{c}
\dot{\textbf{X}}^2 + \acute{\textbf{X}}^2 = \frac{R^2}{R^2 - z^2}\left(\dot{z}^2 + z'^2\right) + \left(R^2 - z^2\right)\left(\dot{\phi}^2 + \phi'^2\right) = R^2 \label{GMVirasoro1} \\[12pt]
\dot{\textbf{X}}\cdot\textbf{X}' = \frac{R^2 \, \dot{z}z'}{R^2 - z^2} + \left(R^2 - z^2\right)\dot{\phi}\phi' = 0 \label{GMVirasoro2} \\[12pt]
\dot{\textbf{X}}^2 - \acute{\textbf{X}}^2 = \frac{R^2}{R^2 - z^2}\left(\dot{z}^2 - z'^2\right) + \left(R^2 - z^2\right)\left(\dot{\phi}^2 - \phi'^2\right) = R^2 \cos2\psi, \label{GMPohlmeyer1}
\end{IEEEeqnarray}
where the Pohlmeyer field $\psi$ satisfies the following sine-Gordon (sG) equation:
\begin{IEEEeqnarray}{ll}
\ddot{\psi} - \psi'' + \frac{1}{2}\sin2\psi = 0. \label{SineGordon1}
\end{IEEEeqnarray}
\indent To obtain single spikes we insert the ansatz
\begin{IEEEeqnarray}{l}
\varphi \equiv \phi - \omega\,\tau = \varphi\left(\sigma - v\omega\tau\right)\,, \quad z = z\left(\sigma - v\omega\tau\right) \label{GM_Ansatz2}
\end{IEEEeqnarray}
into the Virasoro constraints and the Pohlmeyer reduction \eqref{GMVirasoro1}--\eqref{GMPohlmeyer1} and impose the b.c.'s
\begin{IEEEeqnarray}{l}
p \equiv \Delta\phi = \Delta\varphi = \varphi\left(r,\tau\right) - \varphi\left(-r,\tau\right)\,, \quad \Delta z = z\left(r,\tau\right) - z\left(-r,\tau\right) = 0, \label{BoundaryConditions}
\end{IEEEeqnarray}
where $\pm r$ are the string's world-sheet endpoints ($\sigma \in \left[-r,r\right]$) and the string's constant angular extent equals its conserved linear momentum. Equations \eqref{GMVirasoro1}--\eqref{GMPohlmeyer1} become:
\renewcommand{\arraystretch}{2}\begin{table}
\begin{center}\begin{tabular}{|*{4}{c|}}
\cline{1-4}
&$\omega \leq 1$&\multicolumn{2}{|c|}{$\omega \geq 1$}\\
\cline{1-4}
$v\cdot\omega \leq 1$& Doubled Giant Magnon & Elementary Giant Magnon & -- \\[6pt]
\cdashline{1-4}
$v\cdot\omega \geq 1$& -- & Elementary Single Spike & Doubled Single Spike \\[6pt]
\cline{1-4}
&\multicolumn{2}{|c|}{$v \leq 1$}&$v \geq 1$\\[6pt]
\cline{1-4}
\end{tabular}\\\end{center}
\caption{Parameter regions of elementary/doubled giant magnons and single spikes.} \label{Table:GiantMagnons-SingleSpikes}
\end{table}\renewcommand{\arraystretch}{1}
\begin{IEEEeqnarray}{l}
\varphi' = \frac{v\,\omega^2}{1 - v^2\omega^2}\cdot\frac{z^2 - \zeta^2_{\omega}}{R^2 - z^2}\,, \quad \zeta^2_{\omega} \equiv R^2\left[1 - \frac{1}{\omega^2}\right]\,, \quad v\cdot\omega \neq 1 \label{GMVirasoro3} \\[12pt]
z'^2 = \frac{\omega^2}{R^2\left(1 - v^2\omega^2\right)^2}\cdot\left(z^2 - \zeta^2_{\omega}\right)\left(\zeta^2_{v} - z^2\right)\,, \quad \zeta^2_{v} \equiv R^2\left(1 - v^2\right) \label{GMVirasoro4} \\[12pt]
\sin^2\psi = \frac{z^2 - \zeta^2_{\omega}}{\zeta^2_{v} - \zeta^2_{\omega}} \quad \text{(Pohlmeyer reduction)}. \label{GMPohlmeyer2}
\end{IEEEeqnarray}
If $v\cdot\omega = 1$, the system \eqref{GMVirasoro1}--\eqref{GMPohlmeyer1} affords a solution only for $z = 0$ and $v = \omega = 1$. Actually there only exist two solutions, namely the point-like string ($\phi = \pm\tau + \phi_0$) that rotates around the equator of the 2-sphere and its dual under the $\tau\leftrightarrow\sigma$ transform \eqref{TauSigmaTransform} hoop string ($\phi = \pm\sigma + \phi_0$) that is infinitely wrapped around the equator of the 2-sphere and remains at rest. \\[6pt]
\indent One may prove that the constraint equations \eqref{GMVirasoro3}--\eqref{GMVirasoro4} are compatible with the equations of motion that are derived from the action \eqref{GMAction1}, while $\psi$ satisfies the sG equation \eqref{SineGordon1}. Dividing the square root of \eqref{GMVirasoro4} with \eqref{GMVirasoro3}, we obtain:
\begin{IEEEeqnarray}{l}
\frac{dz}{d\varphi} = \frac{R^2 - z^2}{R\,v\,\omega}\sqrt{\frac{\zeta^2_{v} - z^2}{z^2 - \zeta^2_{\omega}}}. \label{GMVirasoro5}
\end{IEEEeqnarray}
\indent All in all, there exist four different solutions of \eqref{GMVirasoro3}--\eqref{GMVirasoro5}, depending on the relative values of the string's linear velocity $v$ and angular velocity $\omega$. These are the following (see table \ref{Table:GiantMagnons-SingleSpikes}):
\begin{IEEEeqnarray}{l}
\text{1. Giant Magnon, Elementary Region (}0 \leq \left|v\right| < 1/\omega \leq 1\text{): }0 \leq \zeta^2_{\omega} = z_{\text{min}}^2 \leq z^2 \leq z_{\text{max}}^2 = \zeta^2_{v} \leq R^2 \nonumber \\[12pt]
\text{2. Giant Magnon, Doubled Region (}0 \leq \left|v\right| \leq 1 \leq 1/\omega\text{): }\zeta^2_{\omega} = - z_{\text{min}}^2 \leq 0 \leq z^2 \leq z_{\text{max}}^2 = \zeta^2_{v} \leq R^2 \nonumber \\[12pt]
\text{3. Single Spike, Elementary Region (}0 \leq 1/\omega < \left|v\right| \leq 1\text{): }0 \leq \zeta^2_{v} = z_{\text{min}}^2 \leq z^2 \leq z_{\text{max}}^2 = \zeta^2_{\omega} \leq R^2 \nonumber \\[12pt]
\text{4. Single Spike, Doubled Region (}0 \leq 1/\omega \leq 1 \leq \left|v\right|\text{): }\zeta^2_{v} = - z_{\text{min}}^2 \leq 0 \leq z^2 \leq z_{\text{max}}^2 = \zeta^2_{\omega} \leq R^2. \nonumber
\end{IEEEeqnarray}
Below we shall focus on the single spike regions (3) and (4), for which $0 \leq 1/\omega \leq \left|v\right| \leq 1$ and $0 \leq 1/\omega \leq 1 \leq \left|v\right|$ respectively. More can be found in appendix A of the paper \cite{FloratosLinardopoulos14}.
%
\begin{figure}
\begin{center}
\includegraphics[scale=0.25]{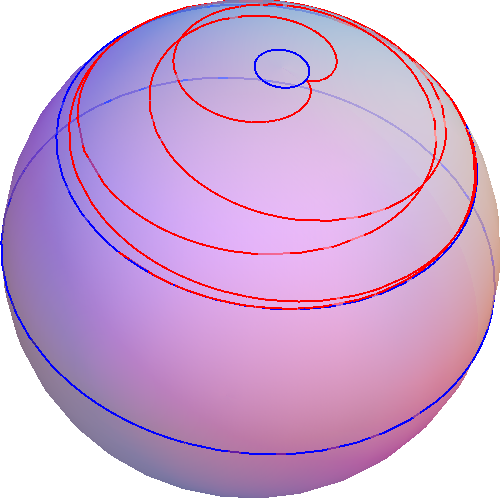} \hspace{2cm} \includegraphics[scale=0.25]{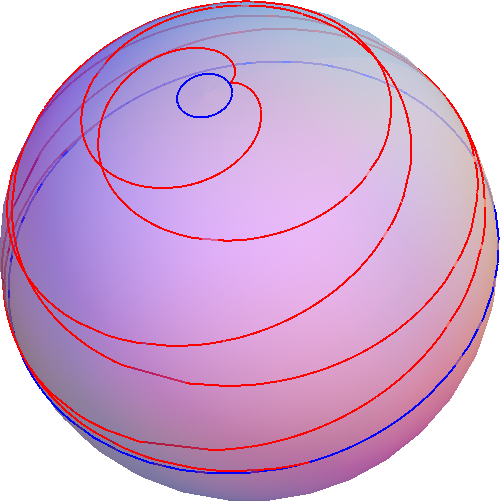}
\caption{Snapshots of elementary (left) and doubled (right) single spike strings ($v\cdot\omega > 1$).} \label{Graph:SingleSpikeStrings}
\end{center}
\end{figure}
%
%
\subsection[Elementary Region]{Single Spikes: Elementary Region}
\begin{figure}
\begin{center}
\includegraphics[scale=0.3]{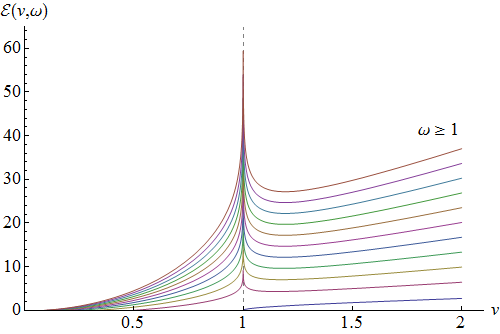}
\
\includegraphics[scale=0.3]{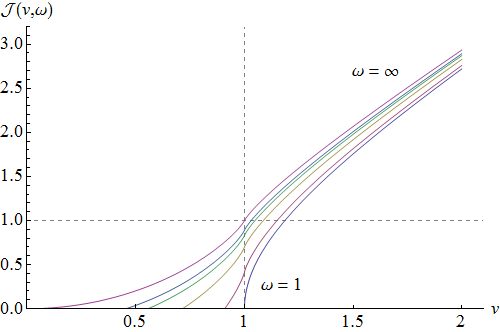}
\
\includegraphics[scale=0.3]{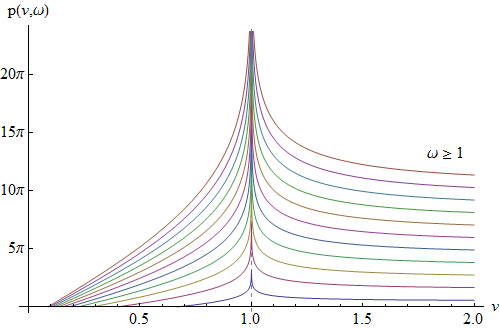}
\caption{Energy, spin and momentum of the single spike as functions of its linear velocity $v$.} \label{Graph:Momentum-Energy-Spin1}
\end{center}
\end{figure}

In the elementary region of single spikes,
\begin{IEEEeqnarray}{c}
0 \leq \frac{1}{\omega} \leq \left|v\right| \leq 1 \qquad \& \qquad 0 \leq \zeta^2_{v} = z_{\text{min}}^2 \leq z^2 \leq z_{\text{max}}^2 = \zeta^2_{\omega} \leq R^2, \label{ElementarySingleSpike}
\end{IEEEeqnarray}
we set:
\begin{IEEEeqnarray}{ll}
\eta \equiv 1 - \frac{z^2_{\text{min}}}{z^2_{\text{max}}} = \frac{v^2 \omega^2 - 1}{\omega^2 - 1} \Leftrightarrow \omega = \sqrt{\frac{1 - \eta}{v^2 - \eta}}. \label{EtaDefinition1}
\end{IEEEeqnarray}
\indent The conserved energy and momenta (angular and linear) of single spikes in the elementary region are given by:\footnote{To obtain the charges of (elementary/doubled) single spikes we must compute $\dot{\phi}$, $\varphi'$ from \eqref{GM_Ansatz2}--\eqref{GMVirasoro3} and $d\sigma$ from \eqref{GMVirasoro4}:
\begin{IEEEeqnarray}{ll}
d\sigma = \frac{dz}{\left|z'\right|} = \frac{R\left(v^2\omega^2 - 1\right)dz}{\omega\sqrt{\left(\zeta^2_{\omega} - z^2\right)\left(z^2 - \zeta^2_{v}\right)}}.
\end{IEEEeqnarray}}
\begin{IEEEeqnarray}{ll}
\mathcal{E} \equiv \frac{\pi E}{\sqrt{\lambda}} = \frac{1}{2} \int_{-r}^{+r} \dot{t} \, d\sigma = r = \frac{v^2 \omega^2 - 1}{\sqrt{\omega^2 - 1}}\cdot\mathbb{K}\left(\eta\right) \label{SS_Energy1} \\[12pt]
\mathcal{J} \equiv \frac{\pi J}{\sqrt{\lambda}} = \frac{1}{2R^2} \int_{-r}^{+r} \left(R^2 - z^2\right)\dot{\phi} \, d\sigma = \sqrt{1 - \frac{1}{\omega^2}} \, \bigg[\mathbb{E}\left(\eta\right) - \frac{1-v^2}{1-1/\omega^2}\,\mathbb{K}\left(\eta\right)\bigg] \label{SS_AngularMomentum1} \\[12pt]
p \equiv \Delta\phi = \Delta\varphi = \int_{-r}^{+r} \varphi' \, d\sigma = \frac{2 v \omega}{\sqrt{1 - 1/\omega^2}} \, \Big[\mathbb{K}\left(\eta\right) - \boldsymbol{\Pi}\left(1 - v^2\omega^2 ; \eta\right)\Big]. \label{SS_Momentum1}
\end{IEEEeqnarray} \\
We have plotted \eqref{SS_Energy1}--\eqref{SS_Momentum1} in terms of the linear velocity $v$ and the angular velocity $\omega$ in figures \ref{Graph:Momentum-Energy-Spin1} and \ref{Graph:Momentum-Energy-Spin2} respectively. The system of equations \eqref{GMVirasoro3}--\eqref{GMPohlmeyer2} admits the following solution: \\
\begin{IEEEeqnarray}{c}
z\left(\tau, \sigma\right) = R\sqrt{1 - \frac{1}{\omega^2}} \cdot \text{dn}\left(\frac{\sigma - v\omega \tau}{\eta\sqrt{\omega^2 - 1}},\eta\right), \quad n \cdot r \leq \sigma - v\omega\tau \leq \left(n + 1\right) \cdot r \label{SingleSpikeZ1}
\end{IEEEeqnarray}
\begin{IEEEeqnarray}{ll}
\varphi\left(z\right) = &\frac{\left(-1\right)^n\,v \omega}{\sqrt{1 - 1/\omega^2}} \Bigg\{\mathbb{F}\bigg(\arcsin\left[\frac{1}{\sqrt{\eta}}\sqrt{1 - \frac{z^2}{z_{\text{max}}^2}}\right],\eta\bigg) - \nonumber \\[12pt]
& \hspace{.5cm} - \boldsymbol{\Pi}\Bigg(1 - v^2\omega^2,\arcsin \left[\frac{1}{\sqrt{\eta}}\sqrt{1 - \frac{z^2}{z_{\text{max}}^2}}\right]\bigg|\,\eta\Bigg)\Bigg\} + \left\lfloor\frac{n + 1}{2}\right\rfloor \cdot p \,, \quad z_{\text{min}} \leq z \leq z_{\text{max}} \qquad \label{SingleSpikePhi1}
\end{IEEEeqnarray}
\begin{IEEEeqnarray}{ll}
\psi\left(\tau, \sigma\right) = \text{am}\left(\frac{\sigma - v\omega \tau}{\eta\,\sqrt{\omega^2 - 1}},\eta\right) \quad \text{(Pohlmeyer reduction)}. \qquad \label{SingleSpikePohlmeyer1}
\end{IEEEeqnarray} \\
Elementary single spikes may be visualized by plotting equation \eqref{SingleSpikePhi1} on a sphere with $\mathsf{Mathematica}$. We obtain the moving string on the left-hand side of figure \ref{Graph:SingleSpikeStrings}.
\newpage
\subsection[Doubled Region]{Single Spikes: Doubled Region}
\begin{figure}
\begin{center}
\includegraphics[scale=0.3]{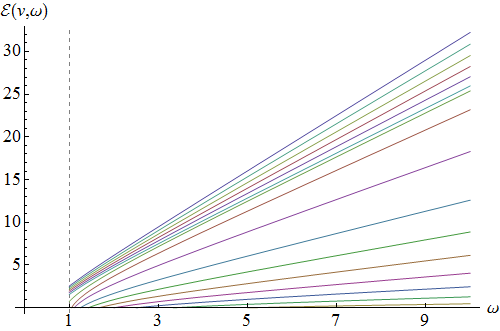}
\
\includegraphics[scale=0.3]{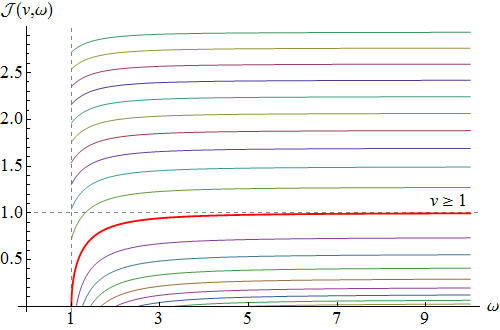}
\
\includegraphics[scale=0.3]{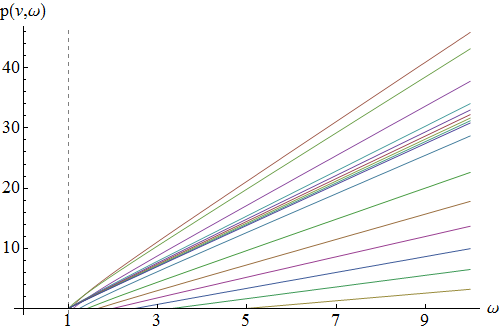}
\caption{Energy, spin and momentum of the single spike as functions of its angular velocity $\omega$.} \label{Graph:Momentum-Energy-Spin2}
\end{center}
\end{figure}
Single spikes in the doubled region have
\begin{IEEEeqnarray}{c}
0 \leq \frac{1}{\omega} \leq 1 \leq \left|v\right| \qquad \& \qquad \zeta^2_{v} = - z_{\text{min}}^2 \leq 0 \leq z^2 \leq z_{\text{max}}^2 = \zeta^2_{\omega} \leq R^2, \label{DoubledSingleSpike}
\end{IEEEeqnarray}
and again we define
\begin{IEEEeqnarray}{ll}
\eta \equiv 1 + \frac{z^2_{\text{min}}}{z^2_{\text{max}}} = \frac{v^2 \omega^2 - 1}{\omega^2 - 1} \Leftrightarrow \omega = \sqrt{\frac{1 - \eta}{v^2 - \eta}}. \label{EtaDefinition2}
\end{IEEEeqnarray} \\
\indent The conserved energy and momenta (angular/linear) of the doubled single spike are given by: \\
\begin{IEEEeqnarray}{ll}
\mathcal{E} = \frac{1}{2} \int_{-r}^{+r} \dot{t} \, d\sigma = r = \sqrt{v^2\omega^2 - 1} \cdot \mathbb{K}\left(\frac{1}{\eta}\right) \label{SS_Energy2} \\[12pt]
\mathcal{J} = \frac{1}{2R^2} \int_{-r}^{+r} \left(R^2 - z^2\right)\dot{\phi} \, d\sigma = \frac{\sqrt{v^2 \omega^2 - 1}}{\omega} \cdot \mathbb{E}\left(\frac{1}{\eta}\right) \label{SS_AngularMomentum2} \\[12pt]
p \equiv \Delta\phi = \Delta\varphi = \int_{-r}^{+r} \varphi' \, d\sigma = \frac{2 v \omega^2}{\sqrt{v^2 \omega^2 - 1}}\left[\mathbb{K}\left(\frac{1}{\eta}\right) - \boldsymbol{\Pi}\left(1 - \omega^2; \frac{1}{\eta}\right)\right]. \label{SS_Momentum2}
\end{IEEEeqnarray} \\
The plots of \eqref{SS_Energy2}--\eqref{SS_Momentum2} in terms of the velocities $v$ and $\omega$ can be found in figures \ref{Graph:Momentum-Energy-Spin1} and \ref{Graph:Momentum-Energy-Spin2}. The system of equations \eqref{GMVirasoro3}--\eqref{GMPohlmeyer2} admits the following solution in the region \eqref{DoubledSingleSpike}: \\
\begin{IEEEeqnarray}{c}
z\left(\tau, \sigma\right) = R\sqrt{1 - \frac{1}{\omega^2}} \cdot \text{cn}\left(\frac{\sigma - v\omega \tau}{\sqrt{v^2 \omega^2 - 1}},\frac{1}{\eta}\right), \quad n \cdot r \leq \sigma - v\omega\tau \leq \left(n + 1\right) \cdot r \label{SingleSpikeZ2}
\end{IEEEeqnarray}
\begin{IEEEeqnarray}{ll}
\varphi\left(z\right) = \frac{\left(-1\right)^n\,v \omega^2}{\sqrt{v^2 \omega^2 - 1}} \Bigg\{\mathbb{F}\left(\arccos\left[\frac{z}{z_{\text{max}}}\right],\frac{1}{\eta}\right) - \boldsymbol{\Pi}\Bigg(1 - \omega^2,\arccos\left[\frac{z}{z_{\text{max}}}\right]\bigg|\,&\frac{1}{\eta}\Bigg)\Bigg\} + 2\left\lfloor\frac{n + 1}{2}\right\rfloor \cdot p\,, \nonumber \\[12pt]
& -z_{\text{max}} \leq z \leq z_{\text{max}} \qquad \label{SingleSpikePhi2}
\end{IEEEeqnarray}
\begin{IEEEeqnarray}{l}
\psi\left(\tau, \sigma\right) = \arcsin\left[\frac{1}{\sqrt{\eta}}\,\text{sn}\left(\frac{\sigma - v\omega \tau}{\sqrt{v^2\omega^2 - 1}},\frac{1}{\eta}\right)\right] \quad \text{(Pohlmeyer reduction)}. \qquad \label{SingleSpikePohlmeyer2}
\end{IEEEeqnarray} \\
The doubled single spike has been plotted on the right-hand side of figure \ref{Graph:SingleSpikeStrings}. The open string gradually unwinds from the north pole and starts winding around the south pole. Then the motion repeats itself.
\subsection[Infinite-Size Limit]{Infinite-Size Limit \label{SubSection:InfiniteSizeLimit}}
In the limit $v = 1$, the energy and the linear momentum of both the elementary and the doubled-region single spikes become infinite. Let us briefly review this limiting case before proceeding to examine yet another infinite-size limit of single spikes. For $v = 1$, the single spike ansatz \eqref{GM_Ansatz1}--\eqref{GM_Ansatz2} becomes:
\begin{IEEEeqnarray}{c}
\Big\{t = \tau, \, \rho = \overline{\theta} = \overline{\phi}_1 = \overline{\phi}_2 = 0\Big\} \times \Big\{\theta = \theta\left(\sigma - \omega\tau\right), \, \phi = \omega\tau + \varphi\left(\sigma - \omega\tau\right), \, \theta_1 = \phi_1 = \phi_2 = 0\Big\}. \hspace{1cm}
\end{IEEEeqnarray}
The corresponding charges can be easily computed, giving rise to the dispersion relation \eqref{SingleSpike1}: \\
\begin{IEEEeqnarray}{ll}
\left.\begin{array}{l}
\mathcal{E} = \sqrt{\omega^2 - 1}\cdot\mathbb{K}\left(1\right) = \infty \\[12pt]
\mathcal{J} = \sqrt{1 - 1/\omega^2} \leq 1 \\[12pt]
p = 2\left[\sqrt{\omega^2 - 1}\cdot\mathbb{K}\left(1\right) - \arcsin\sqrt{1 - 1/\omega^2}\right] = \infty
\end{array}\right\}
\Rightarrow \mathcal{E} - \frac{p}{2} = \arcsin\sqrt{1 - \frac{1}{\omega^2}} = \arcsin\mathcal{J}. \hspace{1cm} \label{SingleSpike4}
\end{IEEEeqnarray} \\[6pt]
The solution to the equations of motion \eqref{GMVirasoro3}--\eqref{GMPohlmeyer2} is:
\begin{IEEEeqnarray}{l}
z\left(\tau, \sigma\right) \equiv R\cos\theta\left(\tau, \sigma\right) = R\,\sqrt{1 - \frac{1}{\omega^2}}\cdot\text{sech}\left(\frac{\sigma - \omega\tau}{\sqrt{\omega^2 - 1}}\right) \\[12pt]
\phi\left(\tau, \sigma\right) = \sigma - \arctan\left[\sqrt{\omega^2 - 1} \tanh\left(\frac{\sigma - \omega\tau}{\sqrt{\omega^2 - 1}}\right)\right] \qquad \\[12pt]
\psi\left(\tau, \sigma\right) = \frac{\pi}{2} - 2\arctan e^{\pm\left(\sigma - \omega\tau\right)/\sqrt{\omega^2 - 1}} = \arcsin\tanh\left[\frac{\sigma - \omega\tau}{\sqrt{\omega^2 - 1}}\right] \quad (\text{Pohlmeyer Reduction}). \hspace{1cm}
\end{IEEEeqnarray}
%
\section[The Omega-Infinity Limit of Single Spikes]{The $\omega = \infty$ Limit of Single Spikes \label{Section:OmegaInfinityLimit}}
We are now ready to explore the $\omega = \infty$ limit of single spikes. Let us first prove that the limit $\omega = \infty$ of the ansatz \eqref{GM_Ansatz1}--\eqref{GM_Ansatz2}, in the single spike region ($v\cdot\omega\geq 1$, $\omega \geq 1$), is an infinite-size limit.
\subsection[Elementary Region]{Elementary Region \label{SubSection:ElementaryRegion}}
\begin{figure}
\begin{center}
\includegraphics[scale=0.25]{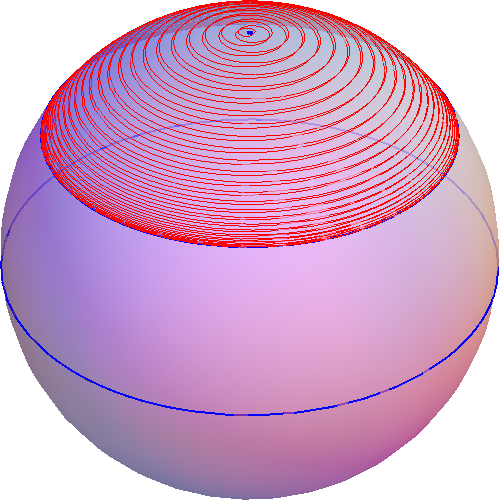} \hspace{1cm} \includegraphics[scale=0.25]{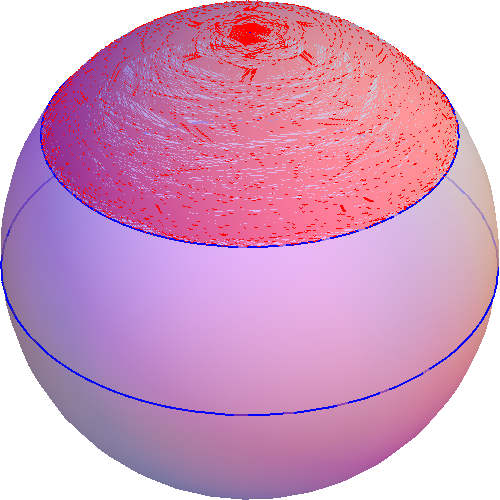} \hspace{1cm} \includegraphics[scale=0.25]{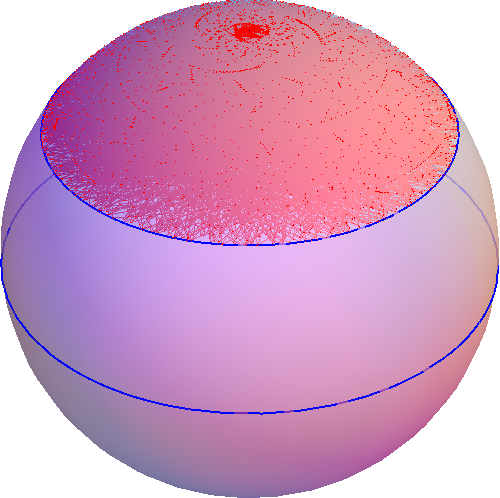}
\caption{Large-$\omega$ elementary ($v = 0.8$) SSs: $\omega = 100$ (left), $\omega = 1000$ (center), $\omega = 10.000$ (right).} \label{Graph:OmegaInfinityElementary}
\end{center}
\end{figure}

We begin with the single spike elementary region for which $v \leq 1$. For $\omega \rightarrow \infty \Leftrightarrow \eta \rightarrow v^2$ the conserved charges \eqref{SS_Energy1}, \eqref{SS_AngularMomentum1}, \eqref{SS_Momentum1} of the elementary single spike become:
\begin{IEEEeqnarray}{ll}
\mathcal{E} = \frac{v^2 \omega^2 - 1}{\sqrt{\omega^2 - 1}}\,\mathbb{K}\left(\eta\right) \rightarrow  v^2 \sqrt{\frac{1 - v^2}{v^2 - \eta}}\cdot \mathbb{K}\left(v^2\right) + O\left(\sqrt{v^2 - \eta}\right), \qquad \eta \equiv \frac{v^2\omega^2 - 1}{\omega^2 - 1} \label{InfiniteSize1} \\[12pt]
\mathcal{J} = \sqrt{1 - \frac{1}{\omega^2}} \, \bigg[\mathbb{E}\left(\eta\right) - \frac{1-v^2}{1-1/\omega^2}\,\mathbb{K}\left(\eta\right)\bigg] \rightarrow \mathbb{E}\left(v^2\right) - \left(1 - v^2\right)\mathbb{K}\left(v^2\right) + O\left(v^2 - \eta\right) \label{InfiniteSize2} \\[12pt]
\frac{p}{2} = \frac{v \omega}{\sqrt{1 - 1/\omega^2}} \, \Big[\mathbb{K}\left(\eta\right) - \boldsymbol{\Pi}\left(1 - v^2\omega^2 ; \eta\right)\Big] \rightarrow v\sqrt{\frac{1 - v^2}{v^2 - \eta}}\cdot\mathbb{K}\left(v^2\right) - \frac{\pi}{2} + O\left(\sqrt{v^2 - \eta}\right). \qquad \label{InfiniteSize3}
\end{IEEEeqnarray} \\
\indent Clearly $\mathcal{E},\, p = \infty$ when $\eta = v^2$. Therefore in the limit $\omega = \infty$, the system's size and its momentum/winding become infinite. See also the plots in figure \ref{Graph:OmegaInfinityElementary}. \\[6pt]
\indent The $\omega = \infty$ dispersion relation of single spikes in the elementary region is the following:
\begin{IEEEeqnarray}{ll}
\mathcal{E} = \left(\frac{p}{2} + \frac{\pi}{2}\right)\cdot v\left(\mathcal{J}\right), \label{InfiniteSizeDispersion1}
\end{IEEEeqnarray}
where $v\left(\mathcal{J}\right)$ gives the string's velocity in terms of its spin. It can be found by inverting \eqref{InfiniteSize2}:
\begin{IEEEeqnarray}{ll}
\mathcal{J} = \mathbb{E}\left(v^2\right) - \left(1 - v^2\right)\mathbb{K}\left(v^2\right), \qquad \omega = \infty. \label{Spin-Velocity1}
\end{IEEEeqnarray}
\indent For $v \rightarrow 0^{+}$, the elliptic integrals do not contain a logarithmic singularity and \eqref{Spin-Velocity1} may easily be inverted with $\mathsf{Mathematica}$. The first few terms of the result are:
\begin{IEEEeqnarray}{l}
v\left(\mathcal{J}\right) = 2\sqrt{\frac{\mathcal{J}}{\pi}}\left[1 - \frac{\mathcal{J}}{4\pi} - \frac{5\mathcal{J}^2}{32\pi^2} - \frac{25\mathcal{J}^3}{128\pi^3} - \frac{653\mathcal{J}^4}{2048\pi^4} - \frac{4935\mathcal{J}^5}{8192\pi^5} + O\left(\mathcal{J}^6\right) \right], \quad \mathcal{J} \rightarrow 0^{+}. \label{InverseVelocity1}
\end{IEEEeqnarray}
\indent For $v \rightarrow 1^{-}$, we may invert \eqref{Spin-Velocity1} by using the series inversion technique that was employed for GKP strings in AdS$_3$ in reference \cite{FloratosGeorgiouLinardopoulos13}. See also appendix \ref{Appendix:InverseSpin}. The result is: \\
\footnotesize\begin{IEEEeqnarray}{ll}
\chi\left(\mathcal{J}\right) = &\chi^* - \frac{2 \mathcal{J} + 3}{64\left(\mathcal{J} - 1\right)}\,\left(\chi^*\right)^3 - \frac{15 \mathcal{J}^2 - 12 \mathcal{J} - 23}{1024\,\left(\mathcal{J} - 1\right)^2}\,\left(\chi^*\right)^4 + \Bigg[\frac{\left(2 \mathcal{J} + 3\right)^2}{8192\left(\mathcal{J} - 1\right)^2} - \frac{351 \mathcal{J}^3 - 964 \mathcal{J}^2 + 209 \mathcal{J} + 644}{49.152\left(\mathcal{J} - 1\right)^3}\Bigg] \left(\chi^*\right)^5 \nonumber \\[6pt]
& + \Bigg[\frac{30\mathcal{J}^3 + 21\mathcal{J}^2 - 82\mathcal{J} - 69}{65.536\left(\mathcal{J} - 1\right)^3} - \frac{1521 \mathcal{J}^4 - 6582 \mathcal{J}^3 + 8282 \mathcal{J}^2 - 370 \mathcal{J} - 3331}{393.216\left(\mathcal{J} - 1\right)^4}\Bigg]\left(\chi^*\right)^6 + \ldots, \ \mathcal{J} \rightarrow 1^-, \ \label{xElementary1}
\end{IEEEeqnarray}\normalsize \\
where $\chi \equiv 1 - v^2$ and $\chi^*$ is given by:
\begin{IEEEeqnarray}{ll}
\chi^* = \frac{4\left(\mathcal{J} - 1\right)}{W_{-1}} = 16\,e^{3/2 - \mathcal{J}/2 + W_{-1}}.
\end{IEEEeqnarray}
The argument of the W-function is $W_{-1}\left[\left(\mathcal{J} - 1\right)e^{\mathcal{J}/2 - 3/2}/4\right]$ in the $W_{-1}$ branch. The $W_{-1}$ branch is the only real branch of the function where $v \rightarrow 1^-$. We may expand the W-function by using its Taylor expansion \eqref{LambertSeries-1} around the point $0^-$. We find,
\small\begin{IEEEeqnarray}{ll}
\chi\left(\mathcal{J}\right) = &\frac{4\mathcal{L}}{\mathcal{S}}\cdot\Bigg\{1 - \left[\ln\mathcal{S} + \frac{\mathcal{L}}{2} + \left(1 + 2\ln2\right)\right]\frac{1}{\mathcal{S}} + \bigg[\ln^2\mathcal{S} + \left[\mathcal{L} + \left(1 + 4\ln2\right)\right]\ln\mathcal{S} - \frac{\mathcal{L}^2}{4} + \left(\frac{7}{4} + 2\ln2\right)\mathcal{L} + \nonumber \\[6pt]
& + 2\ln2\left(1 + 2\ln2\right)\bigg]\frac{1}{\mathcal{S}^2} + \ldots\Bigg\}, \quad \mathcal{L} \equiv 1- \mathcal{J} \rightarrow 0^+, \quad \mathcal{S} \equiv -\ln\left(1 - \mathcal{J}\right) = -\ln\mathcal{L} \rightarrow +\infty, \qquad \label{xElementary2}
\end{IEEEeqnarray}\normalsize
in full agreement with the inverse spin function \eqref{MathematicaInverseSpinFunction1} that we have computed in appendix \ref{Appendix:Computations} with $\mathsf{Mathematica}$. We find the appearance of logarithms in the dispersion relation of strings in $\mathbb{R}\times\text{S}^2$ rather surprising. Usually, the emergence of logarithms in the dispersion relations of single-trace operators of $\mathcal{N} = 4$ SYM is associated with the presence of covariant derivatives inside the operators. Their dual strings then acquire a spin component $S$ inside the non-compact AdS space. As we have also mentioned in the introduction, the inverse spin function of single spikes $\chi\left(\mathcal{J}\right)$ that we find looks very much like the dispersion relation of twist-2 operators Tr$\left[\mathcal{Z}\mathcal{D}_+^{S}\mathcal{Z}\right]$ and the one of their dual closed folded GKP strings in AdS$_3$. However, the characteristic leading logarithm of twist-2 operators is missing from $\chi\left(\mathcal{J}\right)$ and the coefficients of each term depend explicitly on the scaled spin variable $\mathcal{J}$. Yet another difference that could help justify the presence of logarithms in \eqref{xElementary2}, is that the spin variable $\mathcal{S}$ is equal to the logarithm of the real spin $\mathcal{J}$, so that we essentially have reciprocals of logarithms and logarithms of logarithms inside the expression of $\chi\left(\mathcal{J}\right)$ that we must then subtract from unity and take the square root in order to obtain the corresponding dispersion relation. \\[6pt]
\indent In figure \ref{Graph:OmegaInfinityElementary} we have plotted some elementary single spikes for increasing values of the angular velocity $\omega$. The open strings seem to gradually take over all of the space on the 2-sphere between the parallel $z = R\left(1 - v^2\right)^{1/2}$ (that binds their motion from below) and the pole, so that they eventually (as $\omega \rightarrow \infty$) resemble a 2-dimensional brane rather than a string. These "brany strings" are of course very reminiscent of the "stringy branes" that were introduced in \cite{AxenidesFloratosLinardopoulos13a}, i.e.\ M2-branes inside 11-dimensional AdS/CFT spacetimes that have only one stringy mode instead of multiple ones and share the same dispersion relation with strings that live inside 10-dimensional AdS/CFT spacetimes. Of course, we obtain exactly the same graphs by plotting the omega-infinity limits \eqref{SingleSpikeZ3}--\eqref{SingleSpikePhi3} of the solutions \eqref{SingleSpikeZ1}--\eqref{SingleSpikePhi1}. These have been derived in appendix \ref{Appendix:SolutionsCharges}. The sphere-filling property of omega-infinity (elementary) single spikes is proven in appendix \ref{Subsection:SphereFillingProof}.
\subsection[Doubled Region]{Doubled Region \label{SubSection:DoubledRegion}}
\begin{figure}
\begin{center}
\includegraphics[scale=0.25]{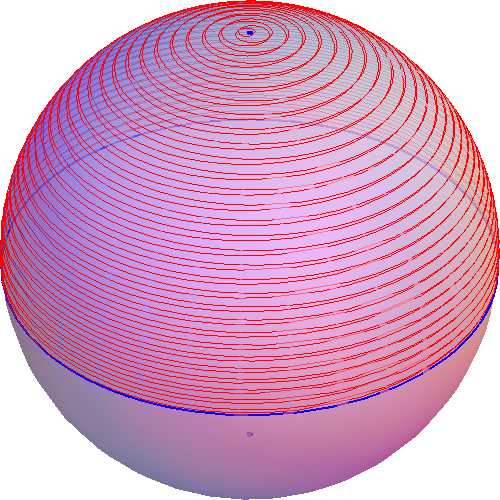} \hspace{1cm} \includegraphics[scale=0.25]{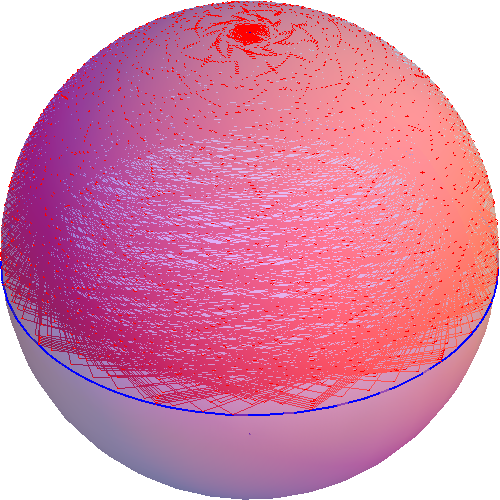} \hspace{1cm} \includegraphics[scale=0.25]{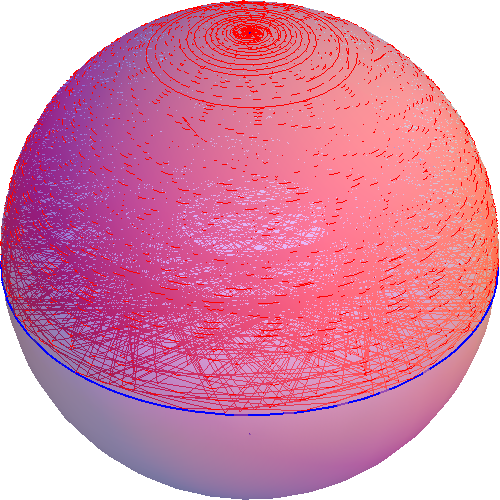}
\caption{Large-$\omega$ doubled ($v = 1.4$) SSs: $\omega = 100$ (left), $\omega = 1000$ (center), $\omega = 20.000$ (right).} \label{Graph:OmegaInfinityDoubled}
\end{center}
\end{figure}

In the doubled region ($v\geq 1$) the conserved charges \eqref{SS_Energy2}, \eqref{SS_AngularMomentum2}, \eqref{SS_Momentum2} of the single spike become, for $\omega \rightarrow \infty \Leftrightarrow \eta \rightarrow v^2$:
\begin{IEEEeqnarray}{ll}
\mathcal{E} = \sqrt{v^2\omega^2 - 1}\cdot\mathbb{K}\left(\frac{1}{\eta}\right) \rightarrow  v \sqrt{\frac{v^2 - 1}{\eta - v^2}}\cdot \mathbb{K}\left(\frac{1}{v^2}\right) + O\left(\sqrt{\eta - v^2}\right), \qquad \eta \equiv \frac{v^2\omega^2 - 1}{\omega^2 - 1} \label{InfiniteSize4} \\[12pt]
\mathcal{J} = \frac{\sqrt{v^2\omega^2 - 1}}{\omega}\cdot\mathbb{E}\left(\frac{1}{\eta}\right) \rightarrow v\,\mathbb{E}\left(\frac{1}{v^2}\right) + O\left(\eta - v^2\right) \label{InfiniteSize5} \\[12pt]
\frac{p}{2} = \frac{v \omega^2}{\sqrt{v^2\omega^2 - 1}} \, \Big[\mathbb{K}\left(\frac{1}{\eta}\right) - \boldsymbol{\Pi}\left(1 - \omega^2 ; \frac{1}{\eta}\right)\Big] \rightarrow \sqrt{\frac{v^2 - 1}{\eta - v^2}}\cdot\mathbb{K}\left(\frac{1}{v^2}\right) - \frac{\pi}{2} + O\left(\sqrt{\eta - v^2}\right). \qquad \label{InfiniteSize6}
\end{IEEEeqnarray} \\
\indent Again $\mathcal{E},p = \infty$ when $\eta = v^2$, and the limit $\omega = \infty$ is an infinite-size/momentum/winding limit. Some plots of doubled single spikes with $\omega \rightarrow \infty$ can be seen in figure \ref{Graph:OmegaInfinityDoubled}. \\[6pt]
\indent The dispersion relation of $\omega = \infty$ single spikes in the doubled region is the following:
\begin{IEEEeqnarray}{ll}
\mathcal{E} = \left(\frac{p}{2} + \frac{\pi}{2}\right)\cdot v\left(\mathcal{J}\right), \label{InfiniteSizeDispersion2}
\end{IEEEeqnarray}
where $v\left(\mathcal{J}\right)$ gives the single spike's velocity in terms of its spin. To find it we must invert \eqref{InfiniteSize5}:
\begin{IEEEeqnarray}{ll}
\mathcal{J} = v\,\mathbb{E}\left(\frac{1}{v^2}\right), \qquad \omega = \infty. \label{Spin-Velocity2}
\end{IEEEeqnarray}
In the limit $v \rightarrow \infty$ the inversion of \eqref{Spin-Velocity2} with $\mathsf{Mathematica}$ is rather straightforward. We find,
\begin{IEEEeqnarray}{l}
v\left(\mathcal{J}\right) = \frac{2\mathcal{J}}{\pi}\left[1 + \frac{\pi^2}{16\mathcal{J}^2} - \frac{\pi^4}{1024\mathcal{J}^4} + \frac{\pi^6}{16.384\mathcal{J}^6} - \frac{13\pi^8}{4.194.304\mathcal{J}^8} + O\left(\frac{1}{\mathcal{J}^{10}}\right)\right], \quad v \rightarrow +\infty. \qquad \label{InverseVelocity2}
\end{IEEEeqnarray} \\
For $v \rightarrow 1^{+}$ we may invert \eqref{InfiniteSizeDispersion1} by the method of reference \cite{FloratosGeorgiouLinardopoulos13} (see also appendix \ref{Appendix:InverseSpin}). The result is:
\footnotesize\begin{IEEEeqnarray}{ll}
&\widetilde{\chi}\left(\mathcal{J}\right) = \widetilde{\chi}^* - \frac{2 \mathcal{J} - 13}{64\left(\mathcal{J} - 1\right)}\left(\widetilde{\chi}^*\right)^3 + \frac{17 \mathcal{J}^2 - 100 \mathcal{J} + 39}{1024\,\left(\mathcal{J} - 1\right)^2}\left(\widetilde{\chi}^*\right)^4 + \Bigg[\frac{\left(2 \mathcal{J} - 13\right)^2}{8192\left(\mathcal{J} - 1\right)^2} - \frac{447 \mathcal{J}^3 - 2468 \mathcal{J}^2 + 625 \mathcal{J} + 868}{49.152\left(\mathcal{J} - 1\right)^3}\Bigg]\left(\widetilde{\chi}^*\right)^5 \nonumber \\[6pt]
& + \Bigg[\frac{34\mathcal{J}^3 - 421\mathcal{J}^2 + 1378\mathcal{J} - 507}{65.536\left(\mathcal{J} - 1\right)^3} - \frac{2103 \mathcal{J}^4 - 11.422 \mathcal{J}^3 + 2030 \mathcal{J}^2 + 10.106 \mathcal{J} - 3853}{393.216\left(\mathcal{J} - 1\right)^4}\Bigg]\left(\widetilde{\chi}^*\right)^6 + \ldots, \ \mathcal{J} \rightarrow 1^+. \ \label{xDoubled1}
\end{IEEEeqnarray}\normalsize \\
This time $\widetilde{\chi} \equiv 1 - 1/v^2$ while $\widetilde{\chi}^*$ is given by:
\begin{IEEEeqnarray}{ll}
\widetilde{\chi}^* = \frac{4\left(1 - \mathcal{J}\right)}{W_{-1}} = 16\,e^{5\mathcal{J}/2 - 5/2 + W_{-1}}.
\end{IEEEeqnarray}
The argument of the W-function is $W_{-1}\left[\left(1 - \mathcal{J}\right)e^{5/2 - 7\mathcal{J}/2}/4\right]$ in the $W_{-1}$ branch. We must choose the $W_{-1}$ branch of the W-function, since it is its only real branch where the $\widetilde{\chi}$ that we find has the desired behavior $\widetilde{\chi} \rightarrow 0^+$ (and therefore $v \rightarrow 1^+$). Again we may expand the W-function by using its Taylor expansion \eqref{LambertSeries-1} around the point $0^-$. Then,
\small\begin{IEEEeqnarray}{ll}
\widetilde{\chi}\left(\mathcal{J}\right) = &\frac{4\mathcal{L}}{\mathcal{S}}\cdot\Bigg\{1 - \left[\ln\mathcal{S} + \frac{7\mathcal{L}}{2} + \left(1 + 2\ln2\right)\right]\frac{1}{\mathcal{S}} + \bigg[\ln^2\mathcal{S} + \left[7\mathcal{L} + \left(1 + 4\ln2\right)\right]\ln\mathcal{S} + \frac{47\mathcal{L}^2}{4} + \left(\frac{25}{4} + 14\ln2\right)\mathcal{L} + \nonumber \\[6pt]
& + 2\ln2\left(1 + 2\ln2\right)\bigg]\frac{1}{\mathcal{S}^2} + \ldots\Bigg\}, \quad \mathcal{L} \equiv \mathcal{J} - 1 \rightarrow 0^+, \quad \mathcal{S} \equiv -\ln\left(\mathcal{J} - 1\right) = -\ln\mathcal{L} \rightarrow +\infty. \qquad \label{xDoubled2}
\end{IEEEeqnarray} \normalsize
\eqref{xDoubled2} fully agrees with the inverse spin function \eqref{MathematicaInverseSpinFunction2} that was computed in appendix \ref{Appendix:Computations} with $\mathsf{Mathematica}$. Once more we note the appearance of logarithms in the dispersion relation. In figure \ref{Graph:OmegaInfinityDoubled} we have plotted some doubled single spikes for increasing values of the angular velocity $\omega$. Exactly the same graphs could have been obtained if we had plotted the omega-infinity limits \eqref{SingleSpikeZ4}--\eqref{SingleSpikePhi4} of the corresponding solutions \eqref{SingleSpikeZ2}--\eqref{SingleSpikePhi2}. See appendix \ref{Subsection:SphereFillingProof} for a proof of the sphere-filling property of doubled single spikes in the omega-infinity limit.
\subsection[General Form of the Dispersion Relation]{General Form of the Dispersion Relation for $\mathcal{J} \rightarrow 1$ \label{SubSection:GeneralForm}}
\begin{figure}
\begin{center}
\includegraphics[scale=0.5]{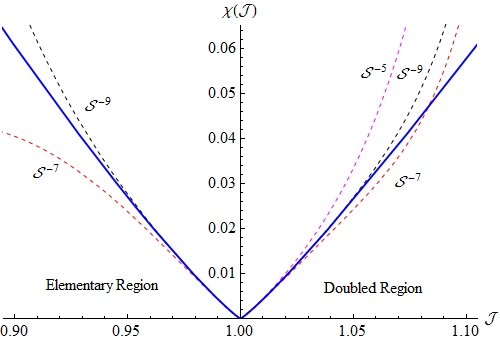}
\caption{Numerical approximations to the inverse spin function $\chi\left(\mathcal{J}\right)$.} \label{Graph:InverseSpinApproximations}
\end{center}
\end{figure}

We are now in a position to write down a general formula for the dispersion relation of $\omega = \infty$ single spikes (both elementary and doubled) in the limit $v\rightarrow 1 \Leftrightarrow \mathcal{J} \rightarrow 1$. The dispersion relation is:
\begin{IEEEeqnarray}{ll}
\mathcal{E} = \left(\frac{p}{2} + \frac{\pi}{2}\right)\cdot v\left(\mathcal{J}\right), \quad \omega = \infty, \quad \lambda = \infty. \label{InfiniteSizeDispersion3}
\end{IEEEeqnarray}
If we define
\begin{IEEEeqnarray}{ll}
v\left(\mathcal{J}\right) = \left[1 - \chi\left(\mathcal{J}\right)\right]^{\pm 1/2} \rightarrow 1^{\mp}, \quad \mathcal{L} \equiv \left|1 - \mathcal{J}\right| \rightarrow 0^+, \quad \mathcal{S} \equiv -\ln\mathcal{L} \rightarrow +\infty,
\end{IEEEeqnarray}
then the inverse spin function $\chi\left(\mathcal{J}\right)$ has the following general form: \\
\begin{IEEEeqnarray}{ll}
\chi\left(\mathcal{J}\right) &= \frac{4\mathcal{L}}{\mathcal{S}}\cdot\left\{\sum_{n = 1}^{\infty}\frac{\ln^n\mathcal{S}}{\mathcal{S}^{n}} + \sum_{n = 2}^{\infty}\rho_{(nn-1)}\frac{\ln^{n-1}\mathcal{S}}{\mathcal{S}^{n}} + \sum_{n = 3}^{\infty}\rho_{(nn-2)}\frac{\ln^{n-2}\mathcal{S}}{\mathcal{S}^{n}} + \ldots + 1 + \sum_{n = 1}^{\infty}\frac{\rho_{n}}{\mathcal{S}^n}\right\} = \nonumber \\[12pt]
&= \frac{4\mathcal{L}}{\mathcal{S}}\cdot\sum_{n = 0}^{\infty}\sum_{k = 0}^{n} \rho_{(nk)}\left(\mathcal{L}\right) \, \frac{\ln^{k}\mathcal{S}}{\mathcal{S}^{n}}, \qquad \rho_{(n0)} \equiv \rho_{n}, \label{InverseSpinFunction2}
\end{IEEEeqnarray} \\
where the $\rho_{(nk)}$'s are given by
\begin{IEEEeqnarray}{ll}
\rho_{(nk)}\left(\mathcal{L}\right) = \sum_{m = 0}^{n - k} \rho_{(nkm)} \cdot \mathcal{L}^{n - k - m}.
\end{IEEEeqnarray}
The values of the coefficients $\rho_{(nkm)}$ (in the elementary and doubled regions respectively) can be read-off from \eqref{xElementary2}--\eqref{xDoubled2}. In figure \ref{Graph:InverseSpinApproximations} we have plotted the spin function $\chi\left(\mathcal{J}\right)$ parametrically, based on \eqref{Spin-Velocity1}--\eqref{Spin-Velocity2} (the precise form is given in equation \eqref{Spin-Velocity4} of appendix \ref{Appendix:InverseSpin}). The dotted lines correspond to the $\mathcal{J} \rightarrow 1$ approximation \eqref{InverseSpinFunction2}, up to and including the $\mathcal{S}^{-5}$, $\mathcal{S}^{-7}$ and $\mathcal{S}^{-9}$ terms.
\section[Discussion]{Discussion \label{Section:Discussion}}
\subsection[Single Spike Dispersion Relations]{Single Spike Dispersion Relations}
Let us summarize what we learned about the dispersion relations of classical single spike strings. The size/momentum/winding of elementary and doubled single spikes becomes infinite in two different limits (see figures \ref{Graph:SingleSpikeRegions1}--\ref{Graph:InfiniteSizeLimits}). The $v = 1$ limit \cite{IshizekiKruczenski07, MosaffaSafarzadeh07} (see \S\ref{SubSection:InfiniteSizeLimit}) in which:
\begin{IEEEeqnarray}{c}
\mathcal{E} - \frac{p}{2} = \arcsin\mathcal{J}, \qquad v = 1, \quad \lambda = \infty \label{SingleSpike5}
\end{IEEEeqnarray}
and the $\omega = \infty$ limit that we studied in \S\ref{Section:OmegaInfinityLimit} of the present paper: \\
\begin{IEEEeqnarray}{ll}
\mathcal{E} = \left(\frac{p}{2} + \frac{\pi}{2}\right) \cdot v\left(\mathcal{J}\right), \qquad \omega = +\infty, \quad \lambda = \infty, \label{SingleSpike6}
\end{IEEEeqnarray} \\
where $v\left(\mathcal{J}\right)$ is found by inverting the spin function \\
\begin{IEEEeqnarray}{ll}
\mathcal{J} = \left\{\begin{array}{ll} \mathbb{E}\left(v^2\right) - \left(1 - v^2\right)\mathbb{K}\left(v^2\right), \quad & 0 \leq v \leq 1 \quad \text{(elementary)} \\[12pt] v\,\mathbb{E}\left(\frac{1}{v^2}\right), & 1 \leq v \leq \infty \quad \text{(doubled).}\end{array}\right.
\end{IEEEeqnarray} \\
\indent For $v \rightarrow 1^{\pm}$, the size/momentum/winding of single spikes is no longer infinite but is still quite large and the dispersion relation \eqref{SingleSpike5} receives exponentially suppressed finite-size corrections that have the following general form (at strong coupling, $\lambda = \infty$) \cite{FloratosLinardopoulos14}:
\begin{IEEEeqnarray}{l}
\mathcal{E} - \frac{p}{2} = \frac{q}{2} + \sum_{n = 1}^{\infty} \bigg[\hat{\mathcal{A}}_{n0} \, p^{2n - 2} + \ldots + \hat{\mathcal{A}}_{n(2n-2)}\bigg] e^{-n\mathcal{R}} = \frac{q}{2} + \sum_{n = 1}^{\infty} \sum_{m = 0}^{2n - 2} \hat{\mathcal{A}}_{nm}\left(q\right) p^{2n - m - 2} e^{-n\mathcal{R}}, \qquad \label{ClassicalCorrections4}
\end{IEEEeqnarray} \\
where $\hat{\mathcal{A}}_{nm}\left(q\right)$ are trigonometric coefficients (see \eqref{ClassicalCorrections2}--\eqref{ClassicalCorrections3}) and we have defined:
\begin{IEEEeqnarray}{ll}
\mathcal{R} \equiv \sqrt{\frac{1}{\mathcal{J}^2} - 1} \cdot \left(p + 2\arcsin\mathcal{J}\right) = \left(p + q\right)\cdot\cot\frac{q}{2}\,, \quad \mathcal{J} \equiv \sin\frac{q}{2}. \label{R-Definition}
\end{IEEEeqnarray}
\begin{figure}
\begin{center}
\includegraphics[scale=0.4]{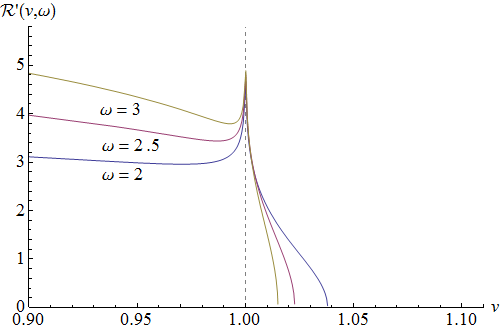} \qquad \includegraphics[scale=0.4]{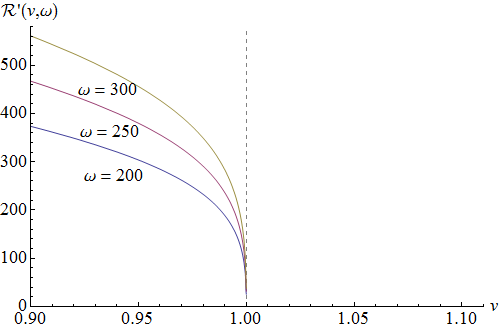}
\caption{Behavior of $\mathcal{R}$ near the limit $v = 1$ for increasing values of the angular velocity $\omega$.} \label{Graph:ExponentialSuppressionCoefficient}
\end{center}
\end{figure}
\indent As long as the exponential suppression factor $e^{-n\mathcal{R}} \rightarrow 0$ for $p \rightarrow \infty$, the dispersion relation \eqref{ClassicalCorrections4} is perfectly well-defined and fully convergent. However it seems that when $\mathcal{J} = 1$, $\mathcal{R}$ becomes finite and the exponential suppression disappears from \eqref{ClassicalCorrections4}. In fact \eqref{ClassicalCorrections4} clearly diverges if we first consider the limit $\mathcal{J} = 1$ and then the limit $p = \infty$. Note also that many of the trigonometric coefficients of \eqref{ClassicalCorrections4} (cf.\ \eqref{ClassicalCorrections2}--\eqref{ClassicalCorrections3}) blow up for $\mathcal{J} = 1 \Leftrightarrow q = \pi$, implying that the value $\mathcal{J} = 1$ of the angular momentum is probably singular. For $\omega = \infty$, it is also possible for $\mathcal{J}$ to approach unity from above ($\mathcal{J} \rightarrow 1^+$---see figure \ref{Graph:InfiniteSizeLimits}), rendering $\mathcal{R}$ in \eqref{R-Definition} a complex number. \\[6pt]
\indent In appendix C of the paper \cite{FloratosLinardopoulos14} it was argued that the dispersion relation \eqref{ClassicalCorrections4} is valid only in the following regime:
\begin{IEEEeqnarray}{ll}
\mathcal{R} \gg 1 \qquad \& \qquad p \gg 1. \label{SingleSpikeConstraint}
\end{IEEEeqnarray}
The reasoning was quite simple: if $\mathcal{R}$ is finite, then the inverse momentum function $x = x\left(p,\mathcal{J}\right)$ (defined in \eqref{x_Parameters} below) that is used in order to obtain \eqref{ClassicalCorrections4} does not have the correct behavior $x \rightarrow 0$ as $v \rightarrow 1$. As a matter of fact we may show that \eqref{SingleSpikeConstraint} becomes less valid as the angular velocity $\omega$ increases towards infinity. Let us define $\mathcal{R}'$ as:
\begin{IEEEeqnarray}{l}
\mathcal{R}' \equiv \sqrt{\frac{1}{\mathcal{J}^2} - 1}\cdot p, \quad p \rightarrow \infty,
\end{IEEEeqnarray}
i.e.\ $\mathcal{R}'$ gives the behavior of the coefficient $\mathcal{R}$ in the large winding regime ($v \rightarrow 1$) where the momentum $p$ becomes infinite. In figure \ref{Graph:ExponentialSuppressionCoefficient} we have plotted $\mathcal{R}'$ near $v = 1$ for various values of the angular velocity $\omega$. There is a delta function peak at $\left\{\omega = \infty,\ v = 1,\ \mathcal{J} = 1\right\}$ but, as we leave the vicinity of this point, $\mathcal{R}$ may become finite and \eqref{ClassicalCorrections4} then is ill-defined and divergent. \\[6pt]
\indent In total there are three paths away from the point $\left\{\omega = \infty,\ v = 1\right\}$ (see figure \ref{Graph:InfiniteSizeLimits}): (a) For $v = 1$ and $\omega \rightarrow \infty$ the dispersion relation of infinite-size/winding single spikes is \eqref{SingleSpike5} which is well-defined and convergent. (b) For $\omega = \infty$ and $v \rightarrow 1^{\pm}$, $\mathcal{R}$ in \eqref{R-Definition} becomes finite and consequently \eqref{SingleSpikeConstraint} and \eqref{ClassicalCorrections4} no longer hold. In this case \eqref{SingleSpike6} is valid and we have shown in \S\ref{SubSection:GeneralForm} that the inverse spin function of single spikes is given by:
\begin{IEEEeqnarray}{ll}
\chi\left(\mathcal{J}\right) &= \frac{4\mathcal{L}}{\mathcal{S}}\cdot\left\{\sum_{n = 1}^{\infty}\frac{\ln^n\mathcal{S}}{\mathcal{S}^{n}} + \sum_{n = 2}^{\infty}\rho_{(nn-1)}\frac{\ln^{n-1}\mathcal{S}}{\mathcal{S}^{n}} + \sum_{n = 3}^{\infty}\rho_{(nn-2)}\frac{\ln^{n-2}\mathcal{S}}{\mathcal{S}^{n}} + \ldots + 1 + \sum_{n = 1}^{\infty}\frac{\rho_{n}}{\mathcal{S}^n}\right\} = \nonumber \\[12pt]
&= \frac{4\mathcal{L}}{\mathcal{S}}\cdot\sum_{n = 0}^{\infty}\sum_{k = 0}^{n} \rho_{(nk)}\left(\mathcal{L}\right) \, \frac{\ln^{k}\mathcal{S}}{\mathcal{S}^{n}}, \qquad \rho_{(n0)} \equiv \rho_{n},
\end{IEEEeqnarray}
where
\begin{IEEEeqnarray}{ll}
\rho_{(nk)}\left(\mathcal{L}\right) = \sum_{m = 0}^{n - k} \rho_{(nkm)} \cdot \mathcal{L}^{n - k - m}.
\end{IEEEeqnarray}
and we have defined \\
\begin{IEEEeqnarray}{ll}
v\left(\mathcal{J}\right) = \left[1 - \chi\left(\mathcal{J}\right)\right]^{\pm 1/2} \rightarrow 1^{\mp}, \quad \mathcal{L} \equiv \left|1 - \mathcal{J}\right| \rightarrow 0^+, \quad \mathcal{S} \equiv -\ln\mathcal{L} \rightarrow +\infty.
\end{IEEEeqnarray} \\
\begin{figure}
\begin{center}
\includegraphics[scale=0.3]{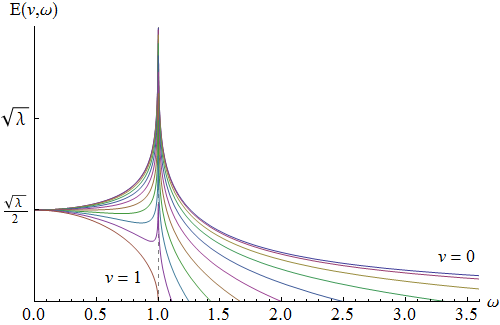}
\
\includegraphics[scale=0.3]{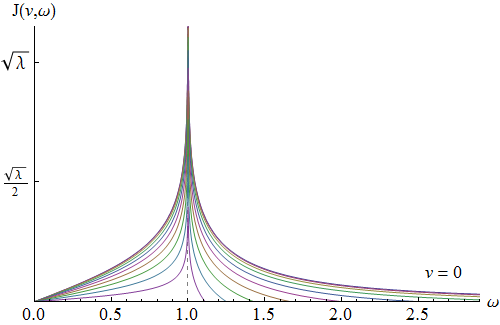}
\
\includegraphics[scale=0.3]{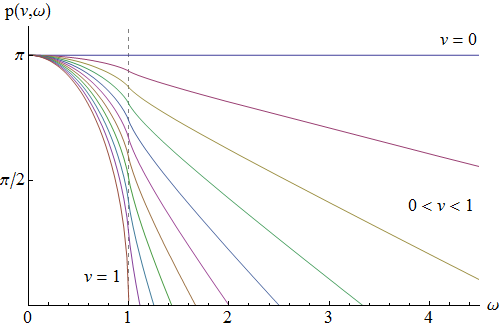}
\caption{Energy, spin and momentum of GMs versus angular velocity. The plot's from \cite{FloratosLinardopoulos14}.} \label{Graph:GiantMagnonCharges}
\end{center}
\end{figure}
Of course, \eqref{SingleSpike5} and \eqref{SingleSpike6} agree with each other in the double infinite-size/momentum/winding limit $\left\{\omega = \infty,\ v = 1,\ \mathcal{J} = 1\right\}$:
\begin{IEEEeqnarray}{c}
\mathcal{E} - \frac{p}{2} = \frac{\pi}{2}, \qquad v = 1, \quad \omega = \infty.
\end{IEEEeqnarray}
(c) Finally, for $\omega \rightarrow \infty, \ v \rightarrow 1^{\pm}$, the dispersion relation \eqref{SingleSpike6} will receive finite-size corrections and its regime of validity is expected to overlap with that of \eqref{ClassicalCorrections4} in certain regions. However there should exist regions where only one of them will be valid.
\subsection[Infinite-Size Limits in $\mathbb{R}\times\text{S}^2$]{Infinite-Size Limits in $\mathbb{R}\times\text{S}^2$}
Here's a short summary of infinitely sized strings in $\mathbb{R}\times\text{S}^2$. Giant magnons acquire infinite size when the angular velocity parameter $\omega$ becomes unity. Single spikes on the other hand become infinitely sized either if their linear velocity becomes unity or if the angular velocity parameter $\omega$ becomes infinite as we have shown in this paper.
\begin{IEEEeqnarray}{ll}
\mathbb{R}\times\text{S}^2\text{ Infinite-Size Limits:} \quad \omega = 1, \quad v = 1, \quad \omega = \infty.
\end{IEEEeqnarray}
All of these infinite-size limits appear in both the elementary and the doubled regions of giant magnons and single spikes. Now it is obvious from the plots of the charges of giant magnons in figure \ref{Graph:GiantMagnonCharges} that $\omega = 1$ is the only infinite-size limit of giant magnons. The corresponding plots of single spikes in figure \ref{Graph:Momentum-Energy-Spin1} however imply the existence of an additional infinite-size limit for them, obtained whenever their velocity $v$ becomes infinite in the doubled region ($0 \leq 1/\omega \leq 1 \leq \left|v\right|$). \\[6pt]
\indent In the $v = \infty$ limit, $1/\eta = 0$ and the charges of doubled single spikes become:
\begin{IEEEeqnarray}{ll}
\mathcal{E} = \frac{\pi}{2}\sqrt{v^2\omega^2 - 1}\rightarrow\infty, \quad
\mathcal{J} = \frac{\pi}{2}\frac{\sqrt{v^2 \omega^2 - 1}}{\omega}\rightarrow\infty, \quad
p = \pi\left(\omega - 1\right) \qquad \left(v = \infty\right). \qquad
\end{IEEEeqnarray}
Therefore the size and the angular momentum of single spikes become infinite for $v = \infty$, whereas their momentum/winding remains a constant. The corresponding dispersion relation is rather simple in this case:
\begin{IEEEeqnarray}{ll}
\mathcal{E} = \left(\frac{p}{\pi} + 1\right)\cdot\mathcal{J} \qquad \left(v = \infty\right).
\end{IEEEeqnarray}
\subsection[Other Topics]{Other Topics}
To avoid possible ambiguities with the limiting procedure $\omega = \infty$, let us give a formal definition of the omega-infinity limit of single spikes. Consider an ordinary single spike with finite values of $\omega > 1$ and $v \neq 1$. Then gradually start increasing $\omega$. The worldsheet size and the physical length of the string will also increase. We identify an omega-infinity single spike with the ideal configuration that results when the size and length of the above string become infinite for $v \neq 1$ and $\omega \rightarrow \infty$. \\[6pt]
\indent The resemblance of \eqref{InverseSpinFunction2} to the generic dispersion relation of twist-2 operators and GKP strings in AdS, that is known to possess the property of reciprocity, makes us want to check whether \eqref{InverseSpinFunction2} also has this remarkable property. By reciprocity we mean that the series
\begin{IEEEeqnarray}{ll}
P\left(\mathcal{S}\right) = \sum_{k = 1}^{\infty} \frac{1}{k!} \left(-\frac{1}{2}\partial_{\mathcal{S}}\right)^{k - 1} \chi^k \left(\mathcal{S}\right) \qquad \label{Reciprocity}
\end{IEEEeqnarray}
of inverse anomalous dimensions contains only even powers of the spin variable $\mathcal{S}$ and it is symmetric under the transformation $\mathcal{S}\rightarrow - \mathcal{S}$. For twist-2 operators and GKP strings, this symmetry leads to a set of relations between the coefficients $\rho_{nk}$ that are known as MVV (Moch-Vermaseren-Vogt) relations. We have explicitly checked \eqref{xElementary2}--\eqref{xDoubled2} for reciprocity and we have found that none of the odd-powered coefficients in the corresponding formulae \eqref{Reciprocity} vanishes. We therefore conclude that the dispersion relation \eqref{InfiniteSizeDispersion3} of single spike strings in the omega-infinity limit, does not have the property of reciprocity. \\[6pt]
\indent Being located near the top of the AdS/CFT spectrum, single spike strings are expected to be unstable. The vacuum (i.e.\ the hoop string) is also unstable as it can roll towards either pole to a configuration of lower energy. The instability of the hoop string and the single spike (elementary/doubled) is most easily proven by using their Pohlmeyer images in the sine-Gordon equation. They correspond to spectrally unstable superluminal ($v\cdot\omega > 1$) rotational/vibrational wave solutions of sG. They may be stabilized however, either by adding more spins \cite{AbbottAniceto08a}, or by turning on appropriate flux terms, or generally by introducing any kind of forcing that could result in a set-up similar to Kapitza's pendulum. In the case of omega-infinity single spikes, we may create a stable closed IIB string in $\mathbb{R}\times\text{S}^2$ by adding a stabilizing counterpart along the opposite hemisphere. \\[6pt]
\indent In appendix \ref{Subsection:SphereFillingProof} we have provided a proof of the sphere-filling property of omega-infinity single spikes. An interesting relevant open question is whether the omega-infinity limit of single spikes can be further shown to satisfy an equation of relativistic membranes on some appropriately chosen spacetime. From the point of view of the Pohlmeyer reduction it would be especially intriguing to know what the reduction of a $2+1$ dimensional surface really looks like, even in the very special case of a brane that is made up from just one very long string. The quantization of these configurations could also lead to very useful conclusions about their true nature, be it stringy or brany. \\[6pt]
\indent Finally, a more or less standard exercise in the context of the present paper, would be to compute the finite-size corrections that are relevant to the $\omega = \infty$ limit of single spikes. The corresponding inverse spin function $x$ has the following form:
\begin{IEEEeqnarray}{l}
x\left(\mathfrak{p},\mathcal{J}\right) = \frac{4\mathcal{L}}{\mathcal{S}}\cdot\sum_{l = 0}^{\infty}\sum_{n = 0}^{\infty}\sum_{k = 0}^{n}r_{(lnk)}\left(\mathcal{L}\right)\,\frac{\ln^{k}\mathcal{S}}{\mathcal{S}^n}\cdot\frac{1}{\mathfrak{p}^{2l}}, \label{FiniteSizeCorrections1}
\end{IEEEeqnarray}
where, in the elementary region ($0 \leq v \leq 1$)
\begin{IEEEeqnarray}{ll}
\mathcal{L} \equiv 1 - \mathcal{J} \rightarrow 0^+, \qquad \mathcal{S} \equiv -\ln\mathcal{L} \rightarrow +\infty, \qquad \mathfrak{p} \equiv p + \pi \rightarrow +\infty
\end{IEEEeqnarray}
and, in order to comply with the notation of \cite{FloratosLinardopoulos14},
\begin{IEEEeqnarray}{ll}
x \equiv 1 - \eta = \frac{1 - v^2}{1 - 1/\omega^2} = \frac{\chi}{1 - 1/\omega^2}, \qquad \chi \equiv 1 - v^2. \label{x_Parameters}
\end{IEEEeqnarray} \\
\indent The first few terms of $x\left(\mathfrak{p},\mathcal{J}\right)$ in the elementary region are the following: \\
\footnotesize\begin{IEEEeqnarray}{ll}
x\left(\mathfrak{p},\mathcal{J}\right) = \frac{4\mathcal{L}}{\mathcal{S}} &- \bigg[4\mathcal{L}\ln\mathcal{S} + 2\mathcal{L}^2 + 4\left(1 + 2\ln2\right)\mathcal{L}\bigg]\frac{1}{\mathcal{S}^2} + \Bigg[4\mathcal{L}\ln^2\mathcal{S} + \left[4\mathcal{L}^2 + 4\left(1 + 4\ln2\right)\mathcal{L}\right]\ln\mathcal{S} - \mathcal{L}^3 + \left(7 + 8 \ln2\right)\mathcal{L}^2 + \nonumber \\[6pt]
& + 8\ln2\left(1 + 2\ln2\right)\mathcal{L}\Bigg]\frac{1}{\mathcal{S}^3} + \ldots + \Bigg\{2(\mathcal{L}-1)\left(\mathcal{S} + \ln\mathcal{S}\right) - 6\mathcal{L}^2 + (6 + 4\ln2)\mathcal{L} - 4\ln2 + \nonumber \\[6pt]
& + \left[2(\mathcal{L}-1)\ln\mathcal{S} - \frac{\mathcal{L}^3}{2} + \frac{7\mathcal{L}^2}{2} - (1 - 4\ln2)\mathcal{L} - 2 - 4\ln2\right]\frac{1}{\mathcal{S}} + \ldots\Bigg\}\cdot\frac{1}{\mathfrak{p}^2} + \ldots, \qquad \label{FiniteSizeCorrections2}
\end{IEEEeqnarray}\normalsize \\
with $\mathfrak{p}$ running in even negative powers. From \eqref{x_Parameters}, we may compute the first few large-$\omega$ corrections to $\chi$: \\
\footnotesize\begin{IEEEeqnarray}{ll}
\chi\left(\mathfrak{p},\mathcal{J}\right) = \frac{4\mathcal{L}}{\mathcal{S}} &- \bigg[4\mathcal{L}\ln\mathcal{S} + 2\mathcal{L}^2 + 4\left(1 + 2\ln2\right)\mathcal{L}\bigg]\frac{1}{\mathcal{S}^2} + \Bigg[4\mathcal{L}\ln^2\mathcal{S} + \left[4\mathcal{L}^2 + 4\left(1 + 4\ln2\right)\mathcal{L}\right]\ln\mathcal{S} - \mathcal{L}^3 + \left(7 + 8 \ln2\right)\mathcal{L}^2 + \nonumber \\[6pt]
& + 8\ln2\left(1 + 2\ln2\right)\mathcal{L}\Bigg]\frac{1}{\mathcal{S}^3} + \ldots - \Bigg\{2(\mathcal{L}+1)\left(\mathcal{S} + \ln\mathcal{S}\right) - 4\mathcal{L}^2 - (10 - 4\ln2)\mathcal{L} + 4\ln2 + \Bigg[2(\mathcal{L}+1)\ln\mathcal{S} - \nonumber \\[6pt]
& - \frac{5\mathcal{L}^3}{2} + \frac{15\mathcal{L}^2}{2} + (9 + 4\ln2)\mathcal{L} + 2 + 4\ln2\Bigg]\frac{1}{\mathcal{S}} + \ldots\Bigg\}\cdot\frac{1}{\mathfrak{p}^2} + \ldots, \qquad \label{FiniteSizeCorrections3}
\end{IEEEeqnarray}\normalsize \\
which are obviously of the form \eqref{FiniteSizeCorrections1}. The corresponding infinite-size coefficients $\rho_{(0nk)}$ for $n =0,\ldots,5$, can be found in appendix \ref{Appendix:Computations}. \\[6pt]
\indent To obtain the large-$\omega$ corrections to the energy of (elementary) single spikes, we must plug the inverse spin function \eqref{FiniteSizeCorrections2} into the corresponding expression of the energy \eqref{SS_Energy1}. We find: \\
\footnotesize\begin{IEEEeqnarray}{ll}
\mathcal{E} &= \Bigg\{\frac{1}{2} - \frac{\mathcal{L}}{\mathcal{S}} + \bigg[\mathcal{L}\ln\mathcal{S} - \frac{\mathcal{L}^2}{2} + (1+2\ln2)\mathcal{L}\bigg]\frac{1}{\mathcal{S}^2} + \ldots\Bigg\}\cdot\mathfrak{p} - \Bigg\{\frac{1}{2}\left(\mathcal{L} + 1\right)\left(\mathcal{S} + \ln\mathcal{S}\right) - \left(\frac{3}{2} - \ln2\right)\mathcal{L} + \ln2 + \nonumber \\[6pt]
& + \left[\frac{1}{2}\left(\mathcal{L} + 1\right)\ln\mathcal{S} - \frac{\mathcal{L}^3}{8} - \frac{13\mathcal{L}^2}{8} + \left(\frac{5}{4} + \ln2\right)\mathcal{L} + \frac{1}{2} + \ln2\right]\frac{1}{\mathcal{S}} + \ldots\Bigg\}\cdot\frac{1}{\mathfrak{p}} + \ldots \qquad \label{FiniteSizeCorrections4}
\end{IEEEeqnarray}\normalsize \\
A simple computation shows that \eqref{FiniteSizeCorrections4} is fully consistent with the infinite-size dispersion relation \eqref{SingleSpike2}--\eqref{xElementary2}. Apparently, the finite-size corrections of the energy run in odd (negative) powers of $\mathfrak{p}$. In the doubled region the results should be similar. Of course, the full-fledged computation of large-omega corrections (in both the elementary and the doubled regions) constitutes a subject matter all by itself. We therefore postpone it for a future publication.
\section[Acknowledgements]{Acknowledgements}
The authors are thankful to an anonymous referee of Nuclear Physics B that brought the issue of convergence of formulas \eqref{ClassicalCorrections2}--\eqref{ClassicalCorrections3} to their attention during the peer-reviewing process of the paper \cite{FloratosLinardopoulos14}. We would like to thank Marius de Leeuw, Benjo Fraser, George Georgiou, Jens Hoppe, Mariusz Hynek, Holger Bech Nielsen, Nikolaos Tetradis and Matthias Wilhelm for illuminating discussions. E.F.\ acknowledges Luis \'{A}lvarez-Gaum\'{e} for instructive discussions and generous support. G.L.\ is grateful to Charlotte Kristjansen and the Niels Bohr Institute for hospitality and support while this work was in progress, and to Jens Hoppe and the KTH Royal Institute of Technology for hospitality and support during the completion of this work. \\[6pt]
\indent The research of E.F.\ has been partially supported by the "ARISTEIA" action (Code no.1612, D.654) and title "Holographic Hydrodynamics" of the "operational programme education and lifelong learning" and is co-funded by the European Social Fund (ESF) and National Resources. The research of M.A.\ and G.L.\ at N.C.S.R.\ "Demokritos" is supported by the General Secretariat for Research and Technology of Greece and from the European Regional Development Fund MIS-448332-ORASY (NSRF 2007--13 ACTION, KRIPIS).
\newpage
\appendix
\section[Solutions \& Charges in the Omega-Infinity Limit]{Solutions \& Charges in the $\omega = \infty$ Limit \label{Appendix:SolutionsCharges}}
In the present appendix we will derive the $\omega = \infty$ solutions of the Virasoro constraints/equations of motion, along with the conserved charges of single spikes. Let us first write down the $\omega = \infty$ limit of the Virasoro constraints \eqref{GMVirasoro3}--\eqref{GMVirasoro4} and the Pohlmeyer reduction \eqref{GMPohlmeyer2}: \\
\begin{IEEEeqnarray}{l}
\varphi' = \frac{1}{v} \Rightarrow \varphi = \frac{\sigma - v\omega\tau}{v}\,, \qquad \eta \equiv \frac{v^2 \omega^2 - 1}{\omega^2 - 1}\rightarrow v^2 \label{GMVirasoro6} \\[12pt]
z'^2 = \frac{1}{R^2\,v^4}\cdot\frac{v^2 - \eta}{\left(1 - v^2\right)}\cdot\left(R^2 - z^2\right)\left(z^2 - \zeta^2_{v}\right)\,, \qquad \zeta^2_{v} \equiv R^2\left(1 - v^2\right) \label{GMVirasoro7} \\[12pt]
\sin^2\psi = \frac{R^2 - z^2}{R^2\,v^2} \quad \text{(Pohlmeyer reduction)}. \label{GMPohlmeyer3}
\end{IEEEeqnarray} \\
We are also going to use the following equations: \\
\begin{IEEEeqnarray}{l}
\frac{d\varphi}{dz} = R\,v\sqrt{\frac{1 - v^2}{\left(v^2 - \eta\right)\left(R^2 - z^2\right)\left(z^2 - \zeta_v^2\right)}} \label{GMEquation1}\\[12pt]
\dot{\phi} = \omega\left(1 - v\,\varphi'\right) = \frac{1}{v^2}\sqrt{\frac{v^2 - \eta}{1 - v^2}}\cdot\frac{z^2 - \zeta_v^2}{R^2 - z^2} \label{GMEquation2}
\end{IEEEeqnarray} \\
in the omega-infinity limit. Notice that although $v^2 - \eta = 0$ for $\omega = \infty$, this factor has explicitly been included in the above equations. The reason is quite subtle. Even though $v^2 - \eta = 0$ for the Virasoro constraint \eqref{GMVirasoro7}, we want to integrate this equation in which case we must take into account the fact that the integrating variable $\xi \equiv \sigma - v\omega\tau$ can become infinite.\footnote{$\sigma$ takes values between $\pm r$, where $r$ is the string's worldsheet size \eqref{InfiniteSize7}--\eqref{InfiniteSize10} (equal to $\infty$ in the omega-infinity limit) and $\tau$ takes values between $\pm\infty$. Therefore $\xi = \sigma - v\omega\tau$ takes values between $\pm\infty$ for single spikes.} This makes $z\left(\sigma - v\omega\tau\right)$ finite and not equal to zero as one could assume by naively taking $\eta = v^2$ in \eqref{GMVirasoro7}. \\[6pt]
\indent The same is true for the conserved charges of $\omega = \infty$ single spikes. The factors $v^2 - \eta$ have to be retained, either because the charge eventually becomes infinite in the $\omega = \infty$ limit or because the corresponding zero/pole cancels and the charge is finite and non-zero. With these considerations in mind we find the following results.
\subsection[Elementary Region]{Elementary Region ($0 \leq 1/\omega \leq \left|v\right| \leq 1$)}
The conserved charges become, for $\omega = \infty \Leftrightarrow \eta = v^2$:
\begin{IEEEeqnarray}{ll}
\mathcal{E} \equiv \frac{\pi E}{\sqrt{\lambda}} = \int_{\zeta_v}^{R} \frac{dz}{\left|z'\right|} = v^2 \sqrt{\frac{1 - v^2}{v^2 - \eta}}\cdot \mathbb{K}\left(v^2\right) = v^2 \sqrt{\omega^2 - 1}\cdot \mathbb{K}\left(v^2\right) \rightarrow \infty \label{InfiniteSize7} \\[12pt]
\mathcal{J} \equiv \frac{\pi J}{\sqrt{\lambda}} = \frac{1}{R^2}\int_{\zeta_v}^{R} \dot{\phi}\left(R^2 - z^2\right)\frac{dz}{\left|z'\right|} = \mathbb{E}\left(v^2\right) - \left(1 - v^2\right)\mathbb{K}\left(v^2\right) \label{InfiniteSize8} \\[12pt]
\frac{p}{2} = \int_{\zeta_v}^{R} \frac{\varphi'\,dz}{\left|z'\right|} = v\,\sqrt{\frac{1 - v^2}{v^2 - \eta}}\cdot\mathbb{K}\left(v^2\right) = v\,\sqrt{\omega^2 - 1}\cdot\mathbb{K}\left(v^2\right) \rightarrow \infty. \qquad \label{InfiniteSize9}
\end{IEEEeqnarray}
%
Note that we retrieve the $\omega = \infty$ dispersion relation $\mathcal{E} = p/2 \cdot v$ (without the $\pi/2$ that is subleading with respect to $p = \infty$). The solutions of the equations of motion are:
\begin{IEEEeqnarray}{ll}
z\left(\tau, \sigma\right) = R \cdot \text{dn}\left(\frac{1}{v^2}\sqrt{\frac{v^2 - \eta}{1 - v^2}}
\cdot \left(\sigma - v\omega \tau\right), v^2\right) = R \cdot \text{dn}&\left(\frac{\sigma - v\omega \tau}{v^2\sqrt{\omega^2 - 1}}, v^2\right) \label{SingleSpikeZ3}
\end{IEEEeqnarray}
\begin{IEEEeqnarray}{ll}
\varphi\left(z\right) = \left(-1\right)^n\,v\,\sqrt{\frac{1 - v^2}{v^2 - \eta}} &\cdot \mathbb{F}\Bigg(\arcsin\left[\frac{1}{v}\sqrt{1 - \frac{z^2}{R^2}}\right], v^2\Bigg) + \left\lfloor\frac{n + 1}{2}\right\rfloor \cdot p = \nonumber \\[6pt]
& = \left(-1\right)^n\,v\,\sqrt{\omega^2 - 1} \cdot \mathbb{F}\left(\arcsin\left[\frac{1}{v}\sqrt{1 - \frac{z^2}{R^2}}\right], v^2\right) + \left\lfloor\frac{n + 1}{2}\right\rfloor \cdot p, \qquad \label{SingleSpikePhi3}
\end{IEEEeqnarray}
where $n \cdot r \leq \sigma - v\omega\tau \leq \left(n + 1\right) \cdot r$. The plots of single spikes in the $\omega = \infty$ limit coincide with the $\omega \rightarrow \infty$ ones that have been drawn in figure \ref{Graph:OmegaInfinityElementary}. The corresponding Pohlmeyer reduction is:
\begin{IEEEeqnarray}{l}
\psi\left(\tau, \sigma\right) = \text{am}\left(\frac{1}{v^2}\sqrt{\frac{v^2 - \eta}{1 - v^2}}
\cdot \left(\sigma - v\omega \tau\right), v^2\right) = \text{am}\left(\frac{\sigma - v\omega \tau}{v^2\sqrt{\omega^2 - 1}}, v^2\right). \qquad \label{SingleSpikePohlmeyer3}
\end{IEEEeqnarray}
\subsection[Doubled Region]{Doubled Region ($0 \leq 1/\omega \leq 1 \leq \left|v\right|$)}
The conserved charges become, for $\omega = \infty \Leftrightarrow \eta = v^2$:
\begin{IEEEeqnarray}{ll}
\mathcal{E} \equiv \frac{\pi E}{\sqrt{\lambda}} = \int_{0}^{R} \frac{dz}{\left|z'\right|} = v \sqrt{\frac{v^2 - 1}{\eta - v^2}} \cdot \mathbb{K}\left(\frac{1}{v^2}\right) = v \sqrt{\omega^2 - 1}\cdot \mathbb{K}\left(\frac{1}{v^2}\right) \rightarrow \infty \label{InfiniteSize10} \\[12pt]
\mathcal{J} \equiv \frac{\pi J}{\sqrt{\lambda}} = \frac{1}{R^2}\int_{0}^{R} \dot{\phi}\left(R^2 - z^2\right)\frac{dz}{\left|z'\right|} = v\,\mathbb{E}\left(\frac{1}{v^2}\right) \label{InfiniteSize11} \\[12pt]
\frac{p}{2} = \int_{0}^{R} \frac{\varphi'\,dz}{\left|z'\right|} = \sqrt{\frac{v^2 - 1}{\eta - v^2}}\cdot\mathbb{K}\left(\frac{1}{v^2}\right) = v\sqrt{\omega^2 - 1}\cdot\mathbb{K}\left(\frac{1}{v^2}\right) \rightarrow \infty. \qquad \label{InfiniteSize12}
\end{IEEEeqnarray}
Again, the $\omega = \infty$ dispersion relation $\mathcal{E} = p/2 \cdot v$ does not contain the subleading contribution $\pi/2$. The equations of motion \eqref{GMVirasoro6}--\eqref{GMVirasoro7} are solved by:
\begin{IEEEeqnarray}{c}
z\left(\tau, \sigma\right) = R \cdot \text{cn}\left(\frac{1}{v}\sqrt{\frac{\eta - v^2}{v^2 - 1}}
\cdot \left(\sigma - v\omega \tau\right), \frac{1}{v^2}\right) = R \cdot \text{cn}\left(\frac{\sigma - v\omega \tau}{v\sqrt{\omega^2 - 1}}, \frac{1}{v^2}\right) \qquad \label{SingleSpikeZ4}
\end{IEEEeqnarray}
\begin{IEEEeqnarray}{ll}
\varphi\left(z\right) = \left(-1\right)^n\sqrt{\frac{v^2 - 1}{\eta - v^2}} \cdot \mathbb{F}\left(\arccos\left[\frac{z}{R}\right], \frac{1}{v^2}\right) &+ \left\lfloor\frac{n + 1}{2}\right\rfloor \cdot p = \nonumber \\[6pt]
&\hspace{-2cm} = \left(-1\right)^n\sqrt{\omega^2 - 1} \cdot \mathbb{F}\left(\arccos\left[\frac{z}{R}\right], \frac{1}{v^2}\right) + \left\lfloor\frac{n + 1}{2}\right\rfloor \cdot p, \quad \label{SingleSpikePhi4}
\end{IEEEeqnarray}
with $n \cdot r \leq \sigma - v\omega\tau \leq \left(n + 1\right) \cdot r$. Plots of single spikes in the $\omega = \infty$ limit can be found in figure \ref{Graph:OmegaInfinityDoubled}. The Pohlmeyer reduction of $\omega = \infty$ single spikes in the doubled region reads:
\begin{IEEEeqnarray}{l}
\psi\left(\tau, \sigma\right) = \arcsin\bigg[\frac{1}{v}\,\text{sn}\Bigg(\frac{1}{v}\sqrt{\frac{\eta - v^2}{v^2 - 1}}
\cdot \left(\sigma - v\omega \tau\right), \frac{1}{v^2}\Bigg)\bigg] = \arcsin\left[\frac{1}{v}\,\text{sn}\left(\frac{\sigma - v\omega \tau}{v\sqrt{\omega^2 - 1}}, \frac{1}{v^2}\right)\right]. \qquad \label{SingleSpikePohlmeyer4}
\end{IEEEeqnarray}
\begin{figure}
\begin{center}
\includegraphics[scale=0.4]{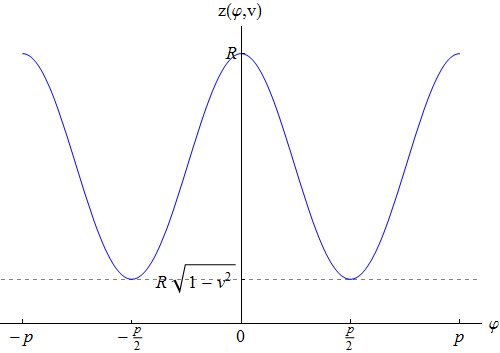} \quad \includegraphics[scale=0.4]{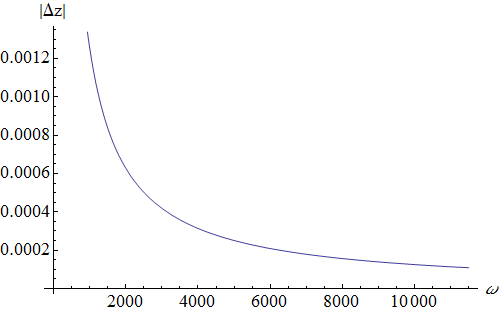}
\caption{$z\left(\varphi,v\right)$ coordinate of omega-infinity single spikes (left) and maximum distance between consecutive string orbits, in terms of the angular velocity $\omega$ (right).} \label{Graph:OrbitDistance}
\end{center}
\end{figure}
\subsection[Proof of the Sphere-Filling Property]{Proof of the Sphere-Filling Property \label{Subsection:SphereFillingProof}}
We may use the solutions \eqref{SingleSpikeZ3}--\eqref{SingleSpikePhi3}/\eqref{SingleSpikeZ4}--\eqref{SingleSpikePhi4} of the (elementary/doubled) single spike Virasoro constraints in the limit $\omega = \infty$ in order to prove that they will eventually become 2-dimensional by completely covering the surface of the 2-sphere between their lowest points and the pole. To this end let us first express the variable $z$ given by \eqref{SingleSpikeZ3}--\eqref{SingleSpikeZ4} (in the limit $\omega = \infty$) in terms of the angle $\varphi$ in \eqref{SingleSpikePhi3}--\eqref{SingleSpikePhi4}: \\
\begin{IEEEeqnarray}{ll}
z\left(\varphi,v\right) = \left\{\begin{array}{ll} R\cdot\text{dn}\left(\varphi/v\sqrt{\omega^2 - 1},v^2\right), \quad & 0 \leq v \leq 1 \quad \text{(elementary)} \\[12pt] R\cdot\text{cn}\left(\varphi/\sqrt{\omega^2 - 1},1/v^2\right), & 1 \leq v \leq \infty \quad \text{(doubled).}\end{array}\right. \label{SingleSpikeOrbits}
\end{IEEEeqnarray} \\
$z\left(\varphi,v\right)$ has been plotted on the left of figure \ref{Graph:OrbitDistance}. It is a periodic function (of period $p$) that oscillates between the parallel $z = R\left(1 - v^2\right)^{1/2}$ and the pole $z = R$. To prove that the string will cover the surface of the 2-sphere in the omega-infinity limit, we need to show that the distance between two adjacent string windings approaches zero as $\omega \rightarrow \infty$. After exactly one winding around the sphere, $z\left(\varphi,v\right)$ becomes $z\left(\varphi + 2\pi,v\right)$. We may plot the function $\Delta z \equiv \left|z\left(\varphi + 2\pi,v\right) - z\left(\varphi,v\right)\right|$\ for increasing angular velocities $\omega$ at the point where it attains its maximum value, namely $\varphi \sim p/4$. The result in both single spike regions (elementary/doubled) is shown on the right of figure \ref{Graph:OrbitDistance}. The same conclusion can be reached if we expand \eqref{SingleSpikeOrbits} around $\omega \rightarrow \infty$:
\begin{IEEEeqnarray}{ll}
\Delta z \equiv \left|z\left(\varphi + 2\pi,v\right) - z\left(\varphi,v\right)\right| \sim \frac{1}{\omega}\sum_{n=0}^{\infty}a_n\left(\frac{\varphi}{\omega}\right)^{2n+1} + O\left(\frac{1}{\omega^2}\right) \rightarrow 0, \quad \text{as} \quad \omega\rightarrow \infty.
\end{IEEEeqnarray}
To complete the proof, note from \eqref{SingleSpikePhi3}--\eqref{SingleSpikePhi4} that $\varphi \sim \omega$, so that $\varphi/\omega \sim$ constant. On the other hand, $z \sim 1 + \varphi^2/\omega^2 + \ldots \neq 0$. Thus two neighboring string orbits come arbitrarily close to each other in the limit $\omega = \infty$ and the string becomes a sphere-filling curve.
%
\section[Inverse Spin]{Inverse Spin \label{Appendix:InverseSpin}}
\begin{figure}[h]
\begin{center}
\includegraphics[scale=0.4]{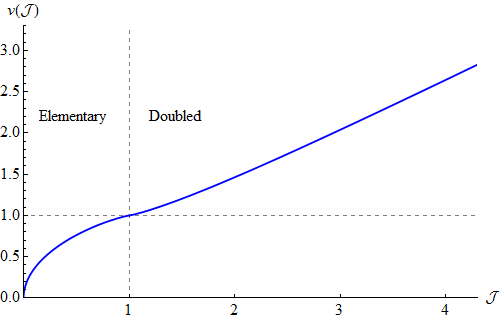} \hspace{1cm} \includegraphics[scale=0.4]{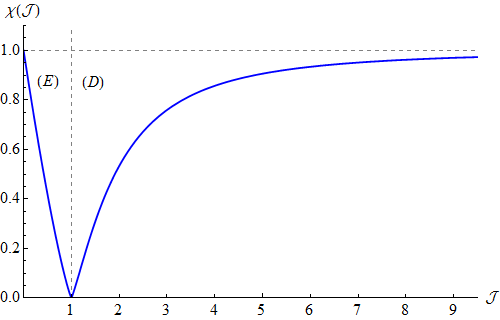}
\caption{Single spike inverse spin functions in the $\omega = \infty$ limit.} \label{Graph:SingleSpikeDispersionRelations}
\end{center}
\end{figure}

This appendix deals with the calculation of the dispersion relation of single spikes in the infinite-size limit $\omega = \infty$, for $v \rightarrow 1^{\pm}$. As we have explained in \S\ref{Section:OmegaInfinityLimit}, the energy of single spikes for $\omega = \infty$ is given by \\
\begin{IEEEeqnarray}{ll}
\mathcal{E} = \left(\frac{p}{2} + \frac{\pi}{2}\right)\cdot v\left(\mathcal{J}\right), \qquad \omega = \infty, \qquad \mathcal{E} \equiv \frac{\pi E}{\sqrt{\lambda}}, \quad \mathcal{J} \equiv \frac{\pi J}{\sqrt{\lambda}}, \label{InfiniteSizeDispersion4}
\end{IEEEeqnarray} \\
where $v\left(\mathcal{J}\right)$ is the expression for the velocity of the single spike in terms of its spin, found by inverting \\
\begin{IEEEeqnarray}{ll}
\mathcal{J} = \left\{\begin{array}{ll} \mathbb{E}\left(v^2\right) - \left(1 - v^2\right)\mathbb{K}\left(v^2\right), \quad & 0 \leq v \leq 1 \quad \text{(elementary)} \\[12pt] v\,\mathbb{E}\left(\frac{1}{v^2}\right), & 1 \leq v \leq \infty \quad \text{(doubled),}\end{array}\right. \label{Spin-Velocity3}
\end{IEEEeqnarray} \\
in terms of the velocity $v$. In the following we are going to show explicitly how \eqref{Spin-Velocity3} can be inverted for $v \rightarrow 1^{\pm}$. Let us first express \eqref{Spin-Velocity3} in terms of the more convenient variables $\chi$ and $\widetilde{\chi}$: \\
\begin{IEEEeqnarray}{ll}
\mathcal{J} = \left\{\begin{array}{ll} \mathbb{E}\left(1 - \chi\right) - \chi\cdot\mathbb{K}\left(1 - \chi\right), \quad & 0 \leq \chi \equiv 1 - v^2 \leq 1 \quad \text{(elementary)} \\[12pt] \left(1 - \widetilde{\chi}\right)^{-1/2}\mathbb{E}\left(1 - \widetilde{\chi}\right), & 0 \leq \widetilde{\chi} \equiv 1 - 1/v^2 \leq 1 \quad \text{(doubled).}\end{array}\right. \label{Spin-Velocity4}
\end{IEEEeqnarray} \\
In figure \ref{Graph:SingleSpikeDispersionRelations}, we have plotted the inverse (spin) functions of \eqref{Spin-Velocity3}--\eqref{Spin-Velocity4}. For simplicity, both variables $\chi$ and $\widetilde{\chi}$ of \eqref{Spin-Velocity4} will henceforth be denoted with $\chi$. \\
 \\
\indent For $\chi \rightarrow 0^{+}$, \eqref{Spin-Velocity4} can be expanded in a series by using the formulas \eqref{EllipticK-Series2}--\eqref{EllipticE-Series2} of appendix \ref{Appendix:EllipticFunctions}:
\begin{IEEEeqnarray}{ll}
\mathcal{J} = \sum_{n = 0}^{\infty} \chi^n \left(c_n\ln \chi + b_n\right), \quad \chi \rightarrow 0^{+}. \label{Spin-Velocity5}
\end{IEEEeqnarray}
In the elementary region, the coefficients $c_n$ and $b_n$ are given by:
\begin{IEEEeqnarray}{c}
c_0 = 0, \quad c_n = \frac{1}{4n}\left(\frac{\left(2n - 3\right)!!}{\left(2n - 2\right)!!}\right)^2, \quad n = 1,2,\ldots \label{SeriesCoefficients1}\\[6pt]
b_0 = 1, \quad b_n = -4c_n\cdot\left\{\ln 2 + H_n - H_{2n} + \frac{2n + 1}{4n\left(2n - 1\right)}\right\}, \quad H_n \equiv \sum_{k=0}^{n}\frac{1}{k}, \quad n = 1,2,\ldots \qquad \label{SeriesCoefficients2}
\end{IEEEeqnarray} \\
$H_n$ are known as harmonic numbers. In the doubled region, the coefficients $c_n$ and $b_n$ are given by:
\begin{IEEEeqnarray}{l}
c_0 = 0, \quad c_n = \sum_{k = 0}^{n} \frac{\left(2k - 1\right)!!}{\left(2k\right)!!}\left(d_{n - k} - f_{n - k}\right), \quad n = 1,2,\ldots \label{SeriesCoefficients3} \\[6pt]
b_0 = 1, \quad b_n = \sum_{k = 0}^{n} \frac{\left(2k - 1\right)!!}{\left(2k\right)!!}\left(h_{n - k} - g_{n - k}\right), \quad n = 1,2,\ldots, \qquad \label{SeriesCoefficients4}
\end{IEEEeqnarray} \\
where,
\begin{IEEEeqnarray}{c}
d_n = -\frac{1}{2}\left(\frac{\left(2n - 1\right)!!}{\left(2n\right)!!}\right)^2, \quad h_n = -4d_n\cdot\left\{\ln 2 + H_n - H_{2n}\right\} \quad n = 1,2,\ldots \label{SeriesCoefficients5} \\[6pt]
f_n = -\frac{d_n}{2n - 1}, \quad g_n = -4f_n\cdot\left\{\ln 2 + H_n - H_{2n} + \frac{1}{2\left(2n - 1\right)}\right\}, \quad n = 1,2,\ldots \qquad \label{SeriesCoefficients6}
\end{IEEEeqnarray} \\
\indent Note that $\mathcal{J} \rightarrow 1^{\pm}$ as $\chi \rightarrow 0^+$, with the minus sign corresponding to the elementary region and the plus sign to the doubled region of single spikes (cf.\ figure \ref{Graph:SingleSpikeDispersionRelations}). The procedure with which \eqref{Spin-Velocity5} can be inverted has been laid down in the paper \cite{FloratosGeorgiouLinardopoulos13}. We repeat it here for convenience. First, we solve \eqref{Spin-Velocity5} in terms of the logarithm:
\begin{IEEEeqnarray}{ll}
\mathcal{J} = b_0 + \sum_{n = 1}^{\infty} \chi^n\Big(c_n\ln \chi + b_n\Big) &\Rightarrow \ln \chi = \left[\mathcal{J} - \sum_{n = 0}^{\infty} b_n \chi^n\right]\cdot\left(\sum_{n = 1}^{\infty} c_n \chi^n\right)^{-1}. \qquad \label{InverseSpinEquation1}
\end{IEEEeqnarray} \\
Secondly we perform all the products and exponentiate the result. This leads us to the following series
\begin{IEEEeqnarray}{ll}
\chi = \chi_0 \cdot \exp\Bigg[\frac{4\left(\mathcal{J} - 1\right)}{\chi} - \frac{1}{16}\left(2\mathcal{J} + 3\right)\chi  - \frac{1}{256}\left(15\mathcal{J} + 11\right)\chi^2 - \ldots\Bigg], \quad \chi_0 \equiv 16\,e^{3/2 - \mathcal{J}/2}, \qquad \label{xEquation1}
\end{IEEEeqnarray} \\
for the inverse spin function in the elementary region ($0 \leq \chi \equiv 1 - v^2 \leq 1$). In the doubled region, the corresponding series becomes (reverting again to the notation $0 \leq \widetilde{\chi} \equiv 1 - 1/v^2 \leq 1$):
\begin{IEEEeqnarray}{ll}
\widetilde{\chi} = \widetilde{\chi}_0 \cdot \exp\Bigg[\frac{4\left(1 - \mathcal{J}\right)}{\widetilde{\chi}} + \frac{1}{16}\left(2\mathcal{J} - 13\right)\widetilde{\chi}  + \frac{1}{256}\left(17\mathcal{J} - 91\right)\widetilde{\chi}^2 - \ldots\Bigg], \quad \widetilde{\chi}_0 \equiv 16\,e^{7\mathcal{J}/2 - 5/2}. \qquad \quad \label{xEquation2}
\end{IEEEeqnarray} \\
Both series \eqref{xEquation1}--\eqref{xEquation2} have the following general form: \\
\begin{IEEEeqnarray}{c}
\chi = \chi_0 \cdot \exp\left[\frac{\text{a}_0}{\chi} + \sum_{n = 1}^{\infty} \text{a}_n \, \chi^n\right] = \chi_0 \cdot \exp\left(\frac{\text{a}_0}{\chi} + \text{a}_1 \, \chi + \text{a}_2 \, \chi^2 + \text{a}_3 \, \chi^3 + \ldots\right), \qquad \label{xEquation3}
\end{IEEEeqnarray} \\
where the $\text{a}_n$'s are linear functions of the (scaled) angular momentum $\mathcal{J} \rightarrow 1^{\pm}$. \eqref{xEquation3} is the same as equation $(6.2)$ of the paper \cite{FloratosGeorgiouLinardopoulos13}. It can be inverted in precisely the same way. Let us repeat the steps here. First, we define the variable $\chi^{*}$ as follows:
\begin{IEEEeqnarray}{c}
\chi^* = \chi_0 \cdot e^{\text{a}_0 / \chi^*} \Rightarrow \chi^* = \frac{\text{a}_0}{W\left(\text{a}_0 / \chi_0\right)} = \chi_0 \cdot e^{W\left(\text{a}_0 / \chi_0\right)}, \label{xstar1}
\end{IEEEeqnarray}
where $W\left(z\right)$ is the Lambert W-function (see appendix \ref{Appendix:LambertFunction}) and $\text{a}_0 \rightarrow 0^{-}$ as $\mathcal{J} \rightarrow 1^{\pm}$. \eqref{xstar1} gives the leading-order solution of \eqref{xEquation3}. From the two real branches of the W-function we have to select the $W_{-1}$ branch (cf.\ figure \ref{Graph:LambertFunction}) for which $W_{-1}\left(\text{a}_0/\chi_0 \rightarrow 0^{-}\right) \rightarrow -\infty$ and $\chi^{*} \rightarrow 0^{+}$.\footnote{Had we chosen the $W_0$ branch, we would have $W_{0}\left(\text{a}_0/\chi_0 \rightarrow 0^{-}\right) \rightarrow 0$ and $\chi^{*} \rightarrow 16e \neq 0$, which is inconsistent with the limit $\chi \rightarrow 0^{+}$ that we are examining.} Further, we set:
\begin{IEEEeqnarray}{c}
\chi = \chi^* \cdot e^u, \label{xstar2}
\end{IEEEeqnarray}
getting the following equation for the variable $u \rightarrow 0$: \\
\begin{IEEEeqnarray}{c}
\left[1 + \frac{\text{a}_0}{\chi^*} - \sum_{k = 1}^{\infty} k \, \text{a}_k \, \left(\chi^*\right)^k\right] \cdot u - \sum_{n = 2}^{\infty} \left[\left(-1\right)^n \frac{\text{a}_0}{\chi^*} + \sum_{k = 1}^{\infty} k^n \, \text{a}_k \, \left(\chi^*\right)^k\right] \frac{u^n}{n!} = \sum_{n = 1}^{\infty} \text{a}_n \, \left(\chi^*\right)^n. \qquad
\end{IEEEeqnarray} \\
This series may be inverted in terms of the variable $u$ and the result may then be plugged into \eqref{xstar2}, giving:
\begin{IEEEeqnarray}{c}
\chi = \chi^* + \frac{\text{a}_1}{\text{a}_0}\left(\chi^*\right)^3 + \left[\frac{\text{a}_2}{\text{a}_0} - \frac{\text{a}_1}{\text{a}_0^2}\right]\left(\chi^*\right)^4 + \left[\frac{\text{a}_3}{\text{a}_0} + \frac{2 \, \text{a}_1^2 - \text{a}_2}{\text{a}_0^2} + \frac{\text{a}_1}{\text{a}_0^3}\right] \left(\chi^*\right)^5 + \ldots, \qquad \label{xGeneralSolution}
\end{IEEEeqnarray}
where,
\begin{IEEEeqnarray}{c}
\chi^* = \frac{\text{a}_0}{W\left(\text{a}_0 / \chi_0\right)} = \chi_0 \cdot e^{W\left(\text{a}_0 / \chi_0\right)}.
\end{IEEEeqnarray}
\indent Plugging the values of the coefficients $\text{a}_n$ (as given by \eqref{xEquation1}--\eqref{xEquation2}) in each of the two regions (the elementary and the doubled) into equation \eqref{xGeneralSolution}, we are led to the expressions \eqref{xElementary1}--\eqref{xDoubled1} that give the inverse spin function of single spike strings with $\omega = \infty$, in the limit $v \rightarrow 1^{\pm}$.
\section[Numerics]{Numerics \label{Appendix:Numerics}}
In this appendix we are going to use $\mathsf{Mathematica}$ in order to compute the inverse spin function $v = v\left(\mathcal{J}\right)$ of $\omega = \infty$ single spikes, in the limit $v\rightarrow 1^{\pm}$. As we have proven in \S\ref{SubSection:ElementaryRegion}--\S\ref{SubSection:DoubledRegion}, the dispersion relation of $\omega = \infty$ single spikes is: \\
\begin{IEEEeqnarray}{ll}
\mathcal{E} = \left(\frac{p}{2} + \frac{\pi}{2}\right)\cdot v\left(\mathcal{J}\right), \qquad \omega = \infty, \qquad \mathcal{E} \equiv \frac{\pi E}{\sqrt{\lambda}}, \quad \mathcal{J} \equiv \frac{\pi J}{\sqrt{\lambda}}. \label{InfiniteSizeDispersion5}
\end{IEEEeqnarray} \\
The inverse spin function $v = v\left(\mathcal{J}\right) \rightarrow 1^{\pm}$ is found by inverting the expressions \eqref{Spin-Velocity1}--\eqref{Spin-Velocity2} for the angular momentum $\mathcal{J} \rightarrow 1^{\pm}$. As we have shown in appendix \ref{Appendix:InverseSpin}, both can essentially be brought to the following form:
\begin{IEEEeqnarray}{ll}
\mathcal{J} = \sum_{n = 0}^{\infty} \chi^n \left(c_n\ln \chi + b_n\right), \quad \chi \rightarrow 0^{+}, \label{Spin-Velocity6}
\end{IEEEeqnarray}
where $\chi \equiv 1 - v^2$ for the elementary region and $\chi \equiv 1 - 1/v^2$ for the doubled region. The coefficients $c_n$ and $b_n$ for each of the two regions have been written down in \eqref{SeriesCoefficients1}--\eqref{SeriesCoefficients6}. \\[6pt]
\indent The method of the paper \cite{FloratosGeorgiouLinardopoulos13}, with which the series \eqref{Spin-Velocity6} can be analytically inverted, has been reviewed in appendix \ref{Appendix:InverseSpin}. The final results \eqref{xElementary1}--\eqref{xDoubled1} can be expressed in terms of the Lambert W-function in its $W_{-1}$ branch. In the present appendix we provide the computer code with which the inverse spin function $\chi = \chi\left(\mathcal{J}\right)$ can be numerically computed with $\mathsf{Mathematica}$ and present the results of the calculations which allow us to directly verify our analytic expressions \eqref{xElementary1}--\eqref{xDoubled1}.
\subsection[Algorithms]{Algorithms \label{Appendix:Algorithms}}
In order to invert equation \eqref{Spin-Velocity6} numerically, we must first cast it in the appropriate form \cite{Linardopoulos15b}. Solving for the logarithm we obtain:
\begin{IEEEeqnarray}{ll}
\ln \chi = \left[\mathcal{J} - \sum_{n = 0}^{\infty} b_n \chi^n\right]\cdot\left(\sum_{n = 1}^{\infty} c_n \chi^n\right)^{-1}, \qquad \label{InverseSpinEquation2}
\end{IEEEeqnarray}
as we did in \eqref{InverseSpinEquation1} above.
\subsubsection[Elementary Region]{Elementary Region}
We now make the following change of variables in \eqref{InverseSpinEquation2}:
\begin{IEEEeqnarray}{ll}
\chi = \frac{4\left(1 - \mathcal{J}\right)}{\mathcal{S}} \cdot e^u, \qquad \mathcal{S} \equiv -\ln\left(1 - \mathcal{J}\right) \label{InverseSpin1}
\end{IEEEeqnarray}
getting, \\
\begin{IEEEeqnarray}{c}
\ln\mathcal{S} = u + 2\ln 2 - \mathcal{S} - \left[\mathcal{J} - \sum_{n = 0}^{\infty} b_n \left[\frac{4\left(1 - \mathcal{J}\right)}{\mathcal{S}} \cdot e^u\right]^n\right]\cdot\left(\sum_{n = 1}^{\infty} c_n \left[\frac{4\left(1 - \mathcal{J}\right)}{\mathcal{S}} \cdot e^u\right]^n\right)^{-1}. \qquad \label{InverseSpin2}
\end{IEEEeqnarray} \\
\indent If we invert this equation for $u$, the variable $\chi = \chi\left(\mathcal{J},\mathcal{S}\right)$ can be obtained from equation \eqref{InverseSpin1}. Here's the $\mathsf{Mathematica}$ code (to run it, just copy-paste it in a $\mathsf{Mathematica}$ notebook):
\begin{Verbatim}[formatcom=\color{blue},baselinestretch=1.5]
c[n_]:=1/(4n)((2n-3)!!/(2n-2)!!)^2;cc[0]:=0;b[0]:=1;
b[n_]:=-4c[n](Log[2]+HarmonicNumber[n]-HarmonicNumber[2n]+(2n+1)/(4n(2n-1)));
y[m_,S_,J_,u_]:=Series[u+2Log[2]-S-(J-Sum[(b[n]((4(1-J)/S)Exp[u])^n),{n,0,m}])
                *(Sum[(c[k]((4(1-J)/S)Exp[u])^k),{k,1,m}])^-1,{u,0,m}];
\[Chi][m_,S_,J_,v_]:=Series[Normal[Series[Normal[(4(1-J))/S
                *Exp[InverseSeries[y[m,S,J,u],v]]],{J,1,m}]],{S,\[Infinity],m}];
nn=5;
\[Chi][nn,S,J,v];
Collect[%/.{v->Log[S],J->1-L},{S,Log[S],Log[2]},FullSimplify]
\end{Verbatim}

\vspace{12pt}
\indent Let us briefly explain what the above code does. ${\color{blue}\tt{c[n]}}$ and ${\color{blue}\tt{b[n]}}$ are just the series coefficients $c_n$, $b_n$ of the elementary region that were given in \eqref{SeriesCoefficients1}--\eqref{SeriesCoefficients2}. Series ${\color{blue}\tt{y[m,S,J,u]}}$ stands for equation \eqref{InverseSpin2} and ${\color{blue}\tt{\backslash[Chi][m,S,J,v]}}$ gives \eqref{InverseSpin1}. The code inverts \eqref{InverseSpin2} for $u$, then plugs the result in \eqref{InverseSpin1} and expands the resulting series around $\mathcal{J} = 1$ and $\mathcal{S} = \infty$. The final output is generated from the last three lines. Instead of $\mathcal{J}$, the variable $\mathcal{L} \equiv 1 - \mathcal{J}$ is used. For ${\color{blue}\tt{nn} = 5}$ terms, the program takes about $15$s to run in our system.
\subsubsection[Doubled Region]{Doubled Region}
In the doubled region, we make the following change of variables in equation \eqref{InverseSpinEquation2}:
\begin{IEEEeqnarray}{ll}
\chi = \frac{4\left(\mathcal{J} - 1\right)}{\mathcal{S}} \cdot e^u, \qquad \mathcal{S} \equiv -\ln\left(\mathcal{J} - 1\right) \label{InverseSpin3}
\end{IEEEeqnarray}
so that \eqref{InverseSpinEquation2} becomes: \\
\begin{IEEEeqnarray}{c}
\ln\mathcal{S} = u + 2\ln 2 - \mathcal{S} - \left[\mathcal{J} - \sum_{n = 0}^{\infty} b_n \left[\frac{4\left(\mathcal{J} - 1\right)}{\mathcal{S}} \cdot e^u\right]^n\right]\cdot\left(\sum_{n = 1}^{\infty} c_n \left[\frac{4\left(\mathcal{J} - 1\right)}{\mathcal{S}} \cdot e^u\right]^n\right)^{-1}. \qquad \label{InverseSpin4}
\end{IEEEeqnarray} \\
\indent Once again, the $\mathsf{Mathematica}$ code consists in inverting \eqref{InverseSpin4} for $u$ and then plugging the result in \eqref{InverseSpin3} in order to obtain the inverse spin function $\chi = \chi\left(\mathcal{J}, \mathcal{S}\right)$ (the code can be directly copy-pasted and run in $\mathsf{Mathematica}$):
\begin{Verbatim}[formatcom=\color{blue},baselinestretch=1.5]
d[n_]:=-(1/2)((2n-1)!!/(2n)!!)^2;
h[n_]:=-4d[n]*(Log[2]+HarmonicNumber[n]-HarmonicNumber[2n]);
f[n_]:=-(d[n]/(2n-1));
g[n_]:=-4f[n]*(Log[2]+HarmonicNumber[n]-HarmonicNumber[2n]+1/(2(2n-1)));
c[n_]:=Sum[((2k-1)!!/(2k)!!)(d[n-k]-f[n-k]),{k,0,n}];
b[n_]:=Sum[((2k-1)!!/(2k)!!)(h[n-k]-g[n-k]),{k,0,n}];
y[m_, S_, J_, u_]:=Series[u+2Log[2]-S-(J-Sum[(b[n]((4(J-1)/S)Exp[u])^n),{n,0,m}])
                   *(Sum[(c[k]((4(J-1)/S)Exp[u])^k),{k,1,m}])^-1,{u,0,m}];
\[Chi][m_, S_, J_, v_]:=Series[Normal[Series[Normal[(4(J-1))/S*
                   Exp[InverseSeries[y[m,S,J,u],v]]],{J,1,m}]],{S,\[Infinity], m}];
nn=5;
\[Chi][nn,S,J,v];
Collect[%/.{v->Log[S],J->L+1},{S,Log[S],Log[2]},FullSimplify]
\end{Verbatim}

\vspace{12pt}
\indent The algorithm is basically the same as the previous one, except for the series coefficients $c_n$, $b_n$, $d_n$, $h_n$, $f_n$, $g_n$ that are given by \eqref{SeriesCoefficients3}--\eqref{SeriesCoefficients6} and are respectively denoted with ${\color{blue}\tt{c[n]}}$, ${\color{blue}\tt{b[n]}}$, ${\color{blue}\tt{d[n]}}$, ${\color{blue}\tt{h[n]}}$, ${\color{blue}\tt{f[n]}}$ and ${\color{blue}\tt{g[n]}}$ in the code above. The series ${\color{blue}\tt{y[m,S,J,u]}}$ and ${\color{blue}\tt{\backslash[Chi][m,S,J,v]}}$ parametrize the equations \eqref{InverseSpin3}--\eqref{InverseSpin4}. The last three lines of the code generate its output. For ${\color{blue}\tt{nn} = 5}$ terms, the program takes about $15$s to run in our system.
\subsection[Computations]{Computations \label{Appendix:Computations}}
Finally, let us give the results for the inverse spin function that we have found by running the above algorithms. They both agree with the inverse spin functions \eqref{xElementary2}--\eqref{xDoubled2} that are calculated from the analytic formulae \eqref{xElementary1}--\eqref{xDoubled1} of \S\ref{SubSection:ElementaryRegion}--\S\ref{SubSection:DoubledRegion} if one replaces the Lambert-W functions with their Taylor expansions \eqref{LambertSeries-1} in the $W_{-1}$ branch.
\subsubsection[Elementary Region]{Elementary Region ($0 \leq v \leq 1$)}
\footnotesize\begin{IEEEeqnarray}{ll}
\chi\left(\mathcal{J}\right) = &\frac{4\mathcal{L}}{\mathcal{S}} - \Bigg[4\mathcal{L}\ln\mathcal{S} + 2\mathcal{L}^2 + 4\left(1 + 2\ln2\right)\mathcal{L}\Bigg]\frac{1}{\mathcal{S}^2} + \Bigg[4\mathcal{L}\ln^2\mathcal{S} + \left[4\mathcal{L}^2 + 4\left(1 + 4\ln2\right)\mathcal{L}\right]\ln\mathcal{S} - \mathcal{L}^3 + \left(7 + 8 \ln2\right)\mathcal{L}^2 + \nonumber \\[6pt]
& + 8\ln2\left(1 + 2\ln2\right)\mathcal{L}\Bigg]\frac{1}{\mathcal{S}^3} - \Bigg[4\mathcal{L}\ln^3\mathcal{S} + \left[6\mathcal{L}^2 + 2\left(1 + 12\ln2\right)\mathcal{L}\right]\ln^2\mathcal{S} - \Big[3\mathcal{L}^3 - \left(17 + 24\ln2\right)\mathcal{L}^2 + \big(4 - 8\ln2 - \nonumber \\[6pt]
& - 48\ln^2 2\big)\mathcal{L}\Big]\ln\mathcal{S} + \frac{5\mathcal{L}^4}{4} - \left(\frac{5}{2} + 6\ln2\right)\mathcal{L}^3 + \left(8 + 34\ln2 + 24\ln^2 2\right)\mathcal{L}^2 - \left(2 + 8\ln2 - 8\ln^2 2 - 32\ln^3 2\right)\mathcal{L}\Bigg]\frac{1}{\mathcal{S}^4} \nonumber \\[6pt]
& + \Bigg[4\mathcal{L}\ln^4\mathcal{S} + \left[8\mathcal{L}^2 - \left(\frac{4}{3} - 32\ln2\right)\mathcal{L}\right]\ln^3\mathcal{S} - \left[6\mathcal{L}^3 - \left(28 + 48\ln2\right)\mathcal{L}^2 + \left(10 + 8\ln2 - 96\ln^2 2\right)\mathcal{L}\right]\ln^2\mathcal{S} + \nonumber \\[6pt]
& + \left[5\mathcal{L}^4 - \left(7 + 24\ln2\right)\mathcal{L}^3 + \left(15 + 112\ln2 + 96\ln^2 2\right)\mathcal{L}^2 - \left(4 + 40\ln2 + 16\ln^2 2 - 128\ln^3 2\right)\mathcal{L}\right]\ln\mathcal{S} - \frac{33\mathcal{L}^5}{16} + \nonumber \\[6pt]
& + \left(\frac{59}{16} + 10\ln2\right)\mathcal{L}^4 + \left(3 - 14\ln2 - 24\ln^2 2\right)\mathcal{L}^3 - \left(2 - 30\ln2 - 112\ln^2 2 - 64\ln^3 2\right)\mathcal{L}^2 + \bigg(\frac{2}{3} - 8\ln2 - \nonumber \\[6pt]
& - 40\ln^2 2 - \frac{32}{3}\ln^3 2 + 64\ln^4 2\bigg)\mathcal{L}\Bigg]\frac{1}{\mathcal{S}^5} + \ldots, \quad \mathcal{L} \equiv 1- \mathcal{J} \rightarrow 0^+, \quad \mathcal{S} \equiv -\ln\left(1 - \mathcal{J}\right) = -\ln\mathcal{L} \rightarrow +\infty. \qquad \label{MathematicaInverseSpinFunction1}
\end{IEEEeqnarray}\normalsize
\subsubsection[Doubled Region]{Doubled Region ($1 \leq v \leq \infty$)}
\footnotesize\begin{IEEEeqnarray}{ll}
\widetilde{\chi}\left(\mathcal{J}\right) = &\frac{4\mathcal{L}}{\mathcal{S}} - \Bigg[4\mathcal{L}\ln\mathcal{S} + 14\mathcal{L}^2 + 4\left(1 + 2\ln2\right)\mathcal{L}\Bigg]\frac{1}{\mathcal{S}^2} + \Bigg[4\mathcal{L}\ln^2\mathcal{S} + \left[28\mathcal{L}^2 + 4\left(1 + 4\ln2\right)\mathcal{L}\right]\ln\mathcal{S} + 47\mathcal{L}^3 + \big(25 + \nonumber \\[6pt]
& + 56\ln2\big)\mathcal{L}^2 + 8\ln2\left(1 + 2\ln2\right)\mathcal{L}\Bigg]\frac{1}{\mathcal{S}^3} - \Bigg[4\mathcal{L}\ln^3\mathcal{S} + \left[42\mathcal{L}^2 + \left(2 + 24\ln2\right)\mathcal{L}\right]\ln^2\mathcal{S} + \big[141\mathcal{L}^3 + \left(47 + 168\ln2\right)\mathcal{L}^2 \nonumber \\[6pt]
& - \left(4 - 8\ln2 - 48\ln^2 2\right)\mathcal{L}\big]\ln\mathcal{S} + \frac{619\mathcal{L}^4}{4} + \left(\frac{235}{2} + 282\ln2\right)\mathcal{L}^3 + \left(8 + 94\ln2 + 168\ln^2 2\right)\mathcal{L}^2 - \big(2 + 8\ln2 - \nonumber \\[6pt]
& - 8\ln^2 2 - 32\ln^3 2\big)\mathcal{L}\Bigg]\frac{1}{\mathcal{S}^4} + \Bigg[4\mathcal{L}\ln^4\mathcal{S} + \left[56\mathcal{L}^2 - \left(\frac{4}{3} - 32\ln2\right)\mathcal{L}\right]\ln^3\mathcal{S} + \Big[282\mathcal{L}^3 + \left(52 + 336\ln2\right)\mathcal{L}^2 - \big(10 + \nonumber \\[6pt]
& + 8\ln2 - 96\ln^2 2\big)\mathcal{L}\Big]\ln^2\mathcal{S} + \Big[619\mathcal{L}^4 + \left(329 + 1128\ln2\right)\mathcal{L}^3 - \left(15 - 208\ln2 - 672\ln^2 2\right)\mathcal{L}^2 - \big(4 + 40\ln2 + \nonumber \\[6pt]
& + 16\ln^2 2 - 128\ln^3 2\big)\mathcal{L}\Big]\ln\mathcal{S} + \frac{8063\mathcal{L}^5}{16} + \left(\frac{7877}{16} + 1238\ln2\right)\mathcal{L}^4 + \left(75 + 658\ln2 + 1128\ln^2 2\right)\mathcal{L}^3 - \big(14 + \nonumber \\[6pt]
& + 30\ln2 - 208\ln^2 2 - 448\ln^3 2\big)\mathcal{L}^2 + \left(\frac{2}{3} - 8\ln2 - 40\ln^2 2 - \frac{32}{3}\ln^3 2 + 64\ln^4 2\right)\mathcal{L}\Bigg]\frac{1}{\mathcal{S}^5} + \ldots, \nonumber \\[6pt]
& \mathcal{L} \equiv \mathcal{J} - 1 \rightarrow 0^+, \quad \mathcal{S} \equiv -\ln\left(\mathcal{J} - 1\right) = -\ln\mathcal{L} \rightarrow +\infty. \qquad \label{MathematicaInverseSpinFunction2}
\end{IEEEeqnarray}\normalsize
\section[Lambert's W-Function]{Lambert's W-Function \label{Appendix:LambertFunction}}
\begin{figure}
\begin{center}
\includegraphics[scale=0.45]{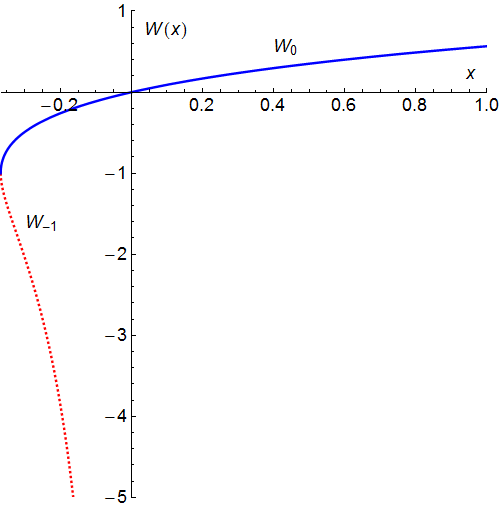}
\caption{Plot of the Lambert W-function.} \label{Graph:LambertFunction}
\end{center}
\end{figure}

In this appendix we repeat some basic properties of the Lambert W-function that we use in our paper. The W-function is defined implicitly as follows:
\begin{IEEEeqnarray}{c}
W\left(z\right)\,e^{W\left(z\right)} = z \Leftrightarrow W\left(z\,e^z\right) = z. \label{LambertDefinition1}
\end{IEEEeqnarray}
The W-function has two real branches, namely $W_0\left(x\right)$ for $x \in \left[-e^{-1},\infty\right)$ and $W_{-1}\left(x\right)$ for $x \in \left[-e^{-1},0\right]$. These are separated by a branch point at $W\left(-e^{-1}\right) = -1$. For a plot of the real branches, see figure \ref{Graph:LambertFunction}. The Taylor expansion of the W-function around the point $x = 0$, in each of the two branches is \cite{CorlessGonnetHareJeffreyKnuth96}:
\begin{IEEEeqnarray}{l}
W_0\left(x\right) = \sum_{n = 0}^\infty \left(-1\right)^n\frac{\left(n+1\right)^n}{\left(n+1\right)!}\cdot x^{n+1} = \sum_{n = 1}^\infty \left(-1\right)^{n + 1} \frac{n^{n - 1}}{n!}\cdot x^n\,, \quad \left|x\right| \leq e^{-1} \label{LambertSeries0} \\[12pt]
W_{-1}\left(x\right) = \ln \left|x\right| - \ln\left|\ln \left|x\right|\right| + \sum_{n = 0}^\infty \sum_{m = 1}^\infty \frac{\left(-1\right)^n}{m!} {n + m \brack n + 1} \left(\ln \left|x\right|\right)^{-n-m} \left(\ln\left|\ln \left|x\right|\right|\right)^m. \qquad \label{LambertSeries-1}
\end{IEEEeqnarray} \\[6pt]
The unsigned Stirling numbers of the first kind $\left[\begin{array}{c}n+m\\n+1\end{array}\right]$ are defined recursively as follows:
\begin{IEEEeqnarray}{c}
\left[\begin{array}{c}n \\ k\end{array}\right] = \left[\begin{array}{c}n - 1 \\ k - 1\end{array}\right] + \left(n - 1\right)\left[\begin{array}{c}n - 1 \\ k\end{array}\right] \quad \& \quad \left[\begin{array}{c}n \\ 0\end{array}\right] = \left[\begin{array}{c}0 \\ k\end{array}\right] = 0\,, \ \left[\begin{array}{c}0 \\ 0\end{array}\right] = 1\,, \quad  n,k\geq 1. \qquad \label{StirlingNumbers1}
\end{IEEEeqnarray}

\section[Elliptic Integrals and Jacobian Elliptic Functions]{Elliptic Integrals and Jacobian Elliptic Functions \label{Appendix:EllipticFunctions}}
This appendix provides the definitions, the Taylor series and some of the properties of elliptic integrals and Jacobian elliptic functions. For details, the reader is referred to the handbook \cite{AbramowitzStegun65}. \\[18pt]
$\blacksquare$ \textbf{Jacobian Elliptic Functions} \\
\begin{IEEEeqnarray}{ll}
u \equiv \int_0^\varphi \frac{d\theta}{\left(1 - m \sin^2\theta\right)^{1/2}} \,, \quad & \varphi \equiv am(u | m) \,, \quad \Delta(\varphi) \equiv (1 - \sin^2\theta)^{1/2} \equiv dn(u | m) \nonumber \\
& x = \sin\varphi \equiv sn(u | m) \,, \quad \cos\varphi \equiv cn(u | m). \nonumber
\end{IEEEeqnarray} \\
$\blacksquare$ \textbf{Elliptic Integral of the First Kind} \\
\begin{IEEEeqnarray}{l}
\mathbb{F}\left(\varphi \big| m\right) \equiv \int_0^\varphi \left(1 - m\,\sin^2\theta\right)^{-1/2} \, d\theta = \int_0^x \left[\left(1 - t^2\right)\left(1 - m\,t^2\right)\right]^{-1/2} \, dt = u \qquad \label{EllipticF1} \\[12pt]
\mathbb{K}\left(m\right) \equiv \mathbb{F}\left(\frac{\pi}{2} \Big| m\right) = \frac{\pi}{2} \cdot {_2\mathcal{F}_1}\left[\frac{1}{2},\frac{1}{2};1;m\right] \quad \text{(complete)} \label{EllipticK1}
\end{IEEEeqnarray}
\begin{IEEEeqnarray}{ll}
\mathbb{K}\left(m\right) = \frac{\pi}{2} \cdot \sum_{n = 0}^\infty &\left(\frac{(2n - 1)!!}{(2n)!!}\right)^2 m^n = \nonumber \\[12pt]
& \qquad = \frac{\pi}{2}\cdot\left[1 + \left(\frac{1}{2}\right)^2 m + \left(\frac{1 \cdot 3}{2 \cdot 4}\right)^2 m^2 + \left(\frac{1 \cdot 3 \cdot 5}{2 \cdot 4 \cdot 6}\right)^2 m^3 + \ldots \right] \,, \quad |m| < 1 \qquad
\end{IEEEeqnarray}
\begin{IEEEeqnarray}{ll}
\mathbb{K}\left(m\right) & = \frac{1}{2 \pi} \cdot \sum_{n = 0}^\infty \left(\frac{\Gamma\left(n + 1/2\right)}{n!}\right)^2 \left[2\psi\left(n + 1\right) - 2\psi\left(n + 1/2\right) - \ln\left(1 - m\right)\right] \, \left(1 - m\right)^n = \nonumber \\[12pt]
& = \sum_{n = 0}^\infty \left(\frac{\left(2n - 1\right)!!}{\left(2n\right)!!}\right)^2 \left[\psi\left(n + 1\right) - \psi\left(n + 1/2\right) - \frac{1}{2}\ln\left(1 - m\right)\right] \, \left(1 - m\right)^n \,, \quad \left|1 - m\right| < 1, \qquad \label{EllipticK-Series2}
\end{IEEEeqnarray} \\[6pt]
where $\psi(z) \equiv \Gamma'(z)/\Gamma(z)$ is the digamma/psi function. \\[18pt]
$\blacksquare$ \textbf{Elliptic Integral of the Second Kind} \\
\begin{IEEEeqnarray}{l}
\mathbb{E}\left(\varphi \big| m\right) \equiv \int_0^\varphi \left(1 - m\,\sin^2\theta\right)^{1/2} \, d\theta = \int_0^x \left(1 - t^2\right)^{-1/2} \left(1 - m\,t^2\right)^{1/2} \, dt \qquad \label{EllipticE1} \\[12pt]
\mathbb{E}\left(m\right) \equiv \mathbb{E}\left(\frac{\pi}{2} \Big| m\right) = \frac{\pi}{2} \cdot {_2\mathcal{F}_1}\left[-\frac{1}{2},\frac{1}{2};1;m\right] \quad \text{(complete)} \label{EllipticE2}
\end{IEEEeqnarray}
\begin{IEEEeqnarray}{ll}
\mathbb{E}\left(m\right) = - \frac{\pi}{2} \cdot \sum_{n = 0}^\infty &\left(\frac{(2n - 1)!!}{(2n)!!}\right)^2 \frac{m^n}{2n - 1} = \nonumber \\[12pt]
& \quad = \frac{\pi}{2} \cdot \left[1 - \left(\frac{1}{2}\right)^2 \frac{m}{1} - \left(\frac{1 \cdot 3}{2 \cdot 4}\right)^2 \frac{m^2}{3} - \left(\frac{1 \cdot 3 \cdot 5}{2 \cdot 4 \cdot 6}\right)^2 \frac{m^3}{5} + \ldots \right] \,, \quad |m| < 1 \qquad
\end{IEEEeqnarray}
\begin{IEEEeqnarray}{lll}
\mathbb{E}\left(m\right) & = 1 &- \frac{1}{2 \pi} \cdot \sum_{n = 0}^\infty \frac{\Gamma\left(n + 1/2\right) \Gamma\left(n + 3/2\right)}{n!\left(n + 1\right)!} \bigg[\ln\left(1 - m\right) + \psi\left(n + 1/2\right) + \psi\left(n + 3/2\right) - \psi\left(n + 1\right) - \nonumber \\[12pt]
&& -\psi\left(n + 2\right)\bigg] \, \left(1 - m\right)^{n + 1} = \nonumber \\[12pt]
& = 1 &+ \sum_{n = 1}^\infty \frac{\left(2n - 1\right)\left[\left(2n - 3\right)!!\right]^2}{\left(2n - 2\right)!!\left(2n\right)!!} \bigg[\psi\left(n\right) - \psi\left(n - 1/2\right) - \frac{1}{2n\left(2n - 1\right)} - \frac{1}{2}\ln\left(1 - m\right)\bigg] \, \left(1 - m\right)^{n} \,, \nonumber \\[12pt]
&& \left|1 - m\right| < 1.\footnote{Note on the double factorial: $0!! = 1$, $\left(-1\right)!! = 1$, $\left(-3\right)!! = -1$.} \qquad \label{EllipticE-Series2}
\end{IEEEeqnarray} \\
$\blacksquare$ \textbf{Elliptic Integral of the Third Kind}
\begin{IEEEeqnarray}{rl}
\boldsymbol{\Pi}(n, \varphi \big| m) &\equiv \int_0^\varphi \left(1 - n \sin^2\theta\right)^{-1} \left(1 - m \sin^2\theta\right)^{-1/2} = \nonumber \\[12pt]
& = \int_0^x \left(1 - n t^2\right)^{-1} \left[\left(1 - t^2\right)\left(1 - m\,t^2\right)\right]^{-1/2} \, dt  \qquad \\[12pt]
\boldsymbol{\Pi}(n;m) &\equiv \boldsymbol{\Pi}(n, \frac{\pi}{2} \Big| m) \quad \text{(complete)}.
\end{IEEEeqnarray} \\
The following addition formula of complete elliptic integrals of the third kind can be found in \cite{ByrdFriedman71}: \\
\begin{IEEEeqnarray}{ll}
\boldsymbol{\Pi}(n;m) = \frac{1}{\left(1-n\right)\mathbb{K}\left(m_1\right)}\cdot\Bigg\{\frac{\pi}{2}&\sqrt{\frac{n\left(n-1\right)}{m-n}}\cdot\mathbb{F}\left(\arcsin\sqrt{\frac{n}{n-m}},m_1\right) - \mathbb{K}\left(m\right)\cdot\bigg[\left(n-1\right)\mathbb{K}\left(m_1\right) - \nonumber \\[12pt]
& - n \cdot \boldsymbol{\Pi}\left(\frac{1-m}{1-n};m_1\right)\bigg]\Bigg\}\,, \quad m + m_1 = 1 \,, \quad 0 < -n < \infty. \qquad \label{AdditionFormula1}
\end{IEEEeqnarray}
\bibliographystyle{JHEP}
\bibliography{Bibliography}

\providecommand{\href}[2]{#2}\begingroup\raggedright\begin{thebibliography}{10}

\bibitem{KristjansenStaudacherTseytlin09}
C.~Kristjansen, M.~Staudacher, and A.~Tseytlin, {\it {Gauge-String Duality and
  Integrability: Progress and Outlook}},  {\em J. Phys.} {\bf A42} (2009)
  250301.

\bibitem{Beisertetal12}
N.~Beisert et~al., {\it {Review of AdS/CFT Integrability: An Overview}},  {\em
  Lett.Math.Phys.} {\bf \textbf{99}} (2012) 3,
  [\href{http://arxiv.org/abs/1012.3982}{{\tt arXiv:1012.3982}}].

\bibitem{MarboeVolin14}
C.~Marboe and D.~Volin, {\it {Quantum Spectral Curve as a Tool for a
  Perturbative Quantum Field Theory}},  {\em Nucl.Phys.} {\bf \textbf{B899}}
  (2015) 810, [\href{http://arxiv.org/abs/1411.4758}{{\tt arXiv:1411.4758}}].

\bibitem{GromovLevkovichMaslyukSizovValatka14}
N.~Gromov, F.~Levkovich-Maslyuk, G.~Sizov, and S.~Valatka, {\it {Quantum
  Spectral Curve at Work: from Small Spin to Strong Coupling in $\mathcal{N} =
  4$ SYM}},  {\em JHEP} {\bf \textbf{07}} (2014) 156,
  [\href{http://arxiv.org/abs/1402.0871}{{\tt arXiv:1402.0871}}].

\bibitem{GromovKazakovLeurentVolin13}
N.~Gromov, V.~Kazakov, S.~Leurent, and D.~Volin, {\it {Quantum Spectral Curve
  for Planar $\mathcal{N} = 4$ Super Yang-Mills Theory}},  {\em Phys.Rev.Lett.}
  {\bf \textbf{112}} (2014) {011602},
  [\href{http://arxiv.org/abs/1305.1939}{{\tt arXiv:1305.1939}}].

\bibitem{Zarembo08talk}
K.~Zarembo, {\it \href{https://www.youtube.com/watch?v=dn6bzwfpTTI}{Spin Chains
  and Strings in Gauge Theory}},  2008.
\newblock Talk at the AlbaNova and Nordita Colloquium.

\bibitem{RoibanTirziuTseytlin06a}
R.~Roiban, A.~Tirziu, and A.~A. Tseytlin, {\it {Slow-String Limit and
  "Antiferromagnetic" State in AdS/CFT}},  {\em Phys.Rev.} {\bf \textbf{D73}}
  (2006) 066003, [\href{http://arxiv.org/abs/hep-th/0601074}{{\tt
  hep-th/0601074}}].

\bibitem{Okamura09}
K.~Okamura, {\it {Giant Spinons}},  {\em JHEP} {\bf \textbf{04}} (2010) 033,
  [\href{http://arxiv.org/abs/0911.1528}{{\tt arXiv:0911.1528}}].

\bibitem{MinahanZarembo03}
J.~A. Minahan and K.~Zarembo, {\it {The Bethe-Ansatz for $\mathcal{N} = 4$
  Super Yang-Mills}},  {\em JHEP} {\bf \textbf{03}} (2003) 013,
  [\href{http://arxiv.org/abs/hep-th/0212208}{{\tt hep-th/0212208}}].

\bibitem{BMN02}
D.~Berenstein, J.~Maldacena, and H.~Nastase, {\it {Strings in Flat Space and pp
  Waves from $\mathcal{N} = 4$ Super Yang Mills}},  {\em JHEP} {\bf
  \textbf{04}} (2002) 013, [\href{http://arxiv.org/abs/hep-th/0202021}{{\tt
  hep-th/0202021}}].

\bibitem{GubserKlebanovPolyakov02}
S.~S. Gubser, I.~R. Klebanov, and A.~M. Polyakov, {\it {A Semi-Classical Limit
  of the Gauge/String Correspondence}},  {\em Nucl.Phys.} {\bf \textbf{B636}}
  (2002) 99, [\href{http://arxiv.org/abs/hep-th/0204051}{{\tt
  hep-th/0204051}}].

\bibitem{HofmanMaldacena06}
D.~M. Hofman and J.~Maldacena, {\it {Giant Magnons}},  {\em J.Phys.} {\bf
  \textbf{A39}} (2006) 13095, [\href{http://arxiv.org/abs/hep-th/0604135}{{\tt
  hep-th/0604135}}].

\bibitem{Zarembo05}
K.~Zarembo, {\it {Antiferromagnetic Operators in $\mathcal{N} = 4$
  Supersymmetric Yang-Mills Theory}},  {\em Phys.Lett.} {\bf \textbf{B634}}
  (2006) 552, [\href{http://arxiv.org/abs/hep-th/0512079}{{\tt
  hep-th/0512079}}].

\bibitem{BeisertDippelStaudacher04}
N.~Beisert, V.~Dippel, and M.~Staudacher, {\it {A Novel Long-Range Spin Chain
  and Planar $\mathcal{N} = 4$ Super Yang-Mills}},  {\em JHEP} {\bf
  \textbf{07}} (2004) 075, [\href{http://arxiv.org/abs/hep-th/0405001}{{\tt
  hep-th/0405001}}].

\bibitem{Beisert05b}
N.~Beisert, {\it {The $\mathfrak{su}\left(2|2\right)$ Dynamic S-Matrix}},  {\em
  Adv.Theor.Math.Phys.} {\bf \textbf{12}} (2008) 945,
  [\href{http://arxiv.org/abs/hep-th/0511082}{{\tt hep-th/0511082}}].

\bibitem{PapathanasiouSpradlin07}
G.~Papathanasiou and M.~Spradlin, {\it {Semiclassical Quantization of the Giant
  Magnon}},  {\em JHEP} {\bf \textbf{06}} (2007) 032,
  [\href{http://arxiv.org/abs/0704.2389}{{\tt arXiv:0704.2389}}].

\bibitem{ChenDoreyLimaMatos07}
H.-Y. Chen, N.~Dorey, and R.~F. Lima~Matos, {\it {Quantum Scattering of Giant
  Magnons}},  {\em JHEP} {\bf \textbf{09}} (2007) 106,
  [\href{http://arxiv.org/abs/0707.0668}{{\tt arXiv:0707.0668}}].

\bibitem{ArutyunovFrolovZamaklar06}
G.~Arutyunov, S.~Frolov, and M.~Zamaklar, {\it {Finite-Size Effects from Giant
  Magnons}},  {\em Nucl.Phys.} {\bf \textbf{B778}} (2007) 1,
  [\href{http://arxiv.org/abs/hep-th/0606126}{{\tt hep-th/0606126}}].

\bibitem{AstolfiForiniGrignaniSemenoff07}
D.~Astolfi, V.~Forini, G.~Grignani, and G.~W. Semenoff, {\it {Gauge Invariant
  Finite Size Spectrum of the Giant Magnon}},  {\em Phys.Lett.} {\bf
  \textbf{B651}} (2007) 329, [\href{http://arxiv.org/abs/hep-th/0702043}{{\tt
  hep-th/0702043}}].

\bibitem{FloratosLinardopoulos14}
E.~Floratos and G.~Linardopoulos, {\it {Large-Spin and Large-Winding Expansions
  of Giant Magnons and Single Spikes}},  {\em Nucl.Phys.} {\bf \textbf{B897}}
  (2015) 229, [\href{http://arxiv.org/abs/1406.0796}{{\tt arXiv:1406.0796}}].

\bibitem{Linardopoulos15b}
G.~Linardopoulos, {\em {Classical Strings and Membranes in the AdS/CFT
  Correspondence}}.
\newblock PhD thesis, National and Kapodistrian University of Athens, 2015.
\newblock \url{http://www.didaktorika.gr/eadd/handle/10442/35838?locale=en}.

\bibitem{Linardopoulos15a}
G.~Linardopoulos, {\it {Large-Spin Expansions of Giant Magnons}},  {\em PoS}
  (CORFU2014) 154, [\href{http://arxiv.org/abs/1502.01630}{{\tt
  arXiv:1502.01630}}].

\bibitem{IshizekiKruczenski07}
R.~Ishizeki and M.~Kruczenski, {\it {Single Spike Solutions for Strings on
  S$^2$ and S$^3$}},  {\em Phys.Rev.} {\bf \textbf{D76}} (2007) 126006,
  [\href{http://arxiv.org/abs/0705.2429}{{\tt arXiv:0705.2429}}].

\bibitem{MosaffaSafarzadeh07}
A.~Mosaffa and B.~Safarzadeh, {\it {Dual Spikes: New Spiky String Solutions}},
  {\em JHEP} {\bf \textbf{08}} (2007) 017,
  [\href{http://arxiv.org/abs/0705.3131}{{\tt arXiv:0705.3131}}].

\bibitem{HayashiOkamuraSuzukiVicedo07}
H.~Hayashi, K.~Okamura, R.~Suzuki, and B.~Vicedo, {\it {Large Winding Sector of
  AdS/CFT}},  {\em JHEP} {\bf \textbf{11}} (2007) 033,
  [\href{http://arxiv.org/abs/0709.4033}{{\tt arXiv:0709.4033}}].

\bibitem{AhnBozhilov08a}
C.~Ahn and P.~Bozhilov, {\it {Finite-Size Effects for Single Spike}},  {\em
  JHEP} {\bf \textbf{07}} (2008) 105,
  [\href{http://arxiv.org/abs/0806.1085}{{\tt arXiv:0806.1085}}].

\bibitem{FloratosGeorgiouLinardopoulos13}
E.~Floratos, G.~Georgiou, and G.~Linardopoulos, {\it {Large-Spin Expansions of
  GKP Strings}},  {\em JHEP} {\bf \textbf{03}} (2014) 018,
  [\href{http://arxiv.org/abs/1311.5800}{{\tt arXiv:1311.5800}}].

\bibitem{AxenidesFloratosLinardopoulos13a}
M.~Axenides, E.~Floratos, and G.~Linardopoulos, {\it {Stringy Membranes in
  AdS/CFT}},  {\em JHEP} {\bf \textbf{08}} (2013) 089,
  [\href{http://arxiv.org/abs/1306.0220}{{\tt arXiv:1306.0220}}].

\bibitem{AbbottAniceto08a}
M.~C. Abbott and I.~V. Aniceto, {\it {Vibrating Giant Spikes and the
  Large-Winding Sector}},  {\em JHEP} {\bf \textbf{06}} (2008) 088,
  [\href{http://arxiv.org/abs/0803.4222}{{\tt arXiv:0803.4222}}].

\bibitem{CorlessGonnetHareJeffreyKnuth96}
R.~M. Corless, G.~H. Gonnet, D.~E.~G. Hare, D.~J. Jeffrey, and D.~E. Knuth,
  {\it {On the Lambert W Function}},  {\em Adv.Comput.Math.} {\bf \textbf{5}}
  (1996) 329.

\bibitem{AbramowitzStegun65}
M.~Abramowitz and I.~Stegun, eds., {\em {Handbook of Mathematical Functions}}.
\newblock Dover, New York, 1972.

\bibitem{ByrdFriedman71}
P.~F. Byrd and M.~D. Friedman, {\em {Handbook of Elliptic Integrals for
  Engineers and Scientists}}.
\newblock Springer-Verlag, 1971.

\end{thebibliography}\endgroup
\end{document}